\documentclass[1p]{elsarticle}
\usepackage{graphicx,amsmath,amssymb,hyperref}

\title{Replicating financial market dynamics with a simple self-organized critical lattice model.}

\author[fiu,fin]{B.~Dupoyet}
\author[fiu,phy]{H.R.~Fiebig}
\author[fiu,phy]{D.P.~Musgrove}

\address[fiu]{Florida International University, Miami, Florida 33199, USA}
\address[fin]{Department of Finance}
\address[phy]{Department of Physics}

\newcommand{\cov}{{\rm cov}}
\newcommand{\var}{{\rm var}}

\begin{document}

\begin{abstract}
We explore a simple lattice field model intended to describe statistical
properties of high frequency financial markets. The model is relevant in the
cross-disciplinary area of econophysics. Its signature feature is the
emergence of a self-organized critical state. This implies scale invariance of
the model, without tuning parameters. Prominent results of our simulation
are time series of gains, prices, volatility, and gains frequency distributions,
which all compare favorably to features of historical market data.
Applying a standard GARCH(1,1) fit to the lattice model gives results that are
almost indistinguishable from historical NASDAQ data. 
\end{abstract}


\maketitle

\section{\label{sec:intro}Introduction}

From a reductionist perspective the statistical physics of a large number of dynamical systems
in nature originates from nonlinear processes at the microscopic level.
In many cases this leads to phenomena characterized in the literature by the terms of chaos,
complexity, fractal geometry, and criticality.
These scenarios are quite ubiquitous, thus not limited to basic physical systems,
where turbulence comes to mind for example, but also to applications in biology,
geology, social networks, economic systems, and finance, to name a few \cite{PerBak:1996}.
The literature on the subject is prodigious.

Our current interest in the subject stems from a recent simulation of financial market
dynamics \cite{Dupoyet2010107}. At the root of that study is a microscopic model based
on the principal of gauge invariance, assuming that one of the key mechanisms of trader
behavior is independent of any scale (currency unit, for example) used in the market
transactions \cite{Ilinski:2001:PF}.
In technical terms the model is a quantum field theory based on the
gauge group $G={\mathbb R}^+$, the dilation group,
which implies scale invariance of the market model with respect to ordinary
multiplication of prices with positive real numbers.
The quantum aspect of that model implements the empirical observation that arbitrage
opportunities, i.e. realizing a profit via transactions in different markets, vanish
quickly because of market dynamics.

At this stage, the model does not provide a mechanism for describing a complex
system, as it should, given the empirical evidence.
The distribution of market returns, if analyzed appropriately \cite{Dupoyet2010107},
exhibits fat tails (probabilities larger than Gaussian)
the likes of which are observed in many high frequency financial markets.
However, this is not a consequence of the intrinsic dynamics of the model.
In order to remedy this situation, in the present article,
we study an abridged model with local interactions that lead to a
self-organized complex market model.
Although our goal is to eventually combine the
gauge model with features of the abridged model discussed here, the latter, despite its
simplicity, produces salient characteristics of actual financial markets surprisingly well.
Among those are the semblance of return time series, returns frequency distributions,
and the nature of volatility. The market volatility, in particular,
is a subject of intense research \cite{Poon:2003,Ryden:1998,Granger:1995}.

These features are promising enough to study this simple model in its own right,
which is the subject of this work. Along a one-dimensional lattice representing
discrete time, the field on the sites are interpreted as returns in a model market.
An updating algorithm is then applied which is loosely fashioned after the well-known
proposal by Bak, Tang and Wiesenfeld (BTW) \cite{PhysRevLett.59.381},
see also \cite{PhysRevA.38.364, PhysRevLett.71.4083, PerBak:1996, PhysRevE.53.414}.
The idea is to develop a market model that is driven by microscopic entities, say
traders, such that their interaction leaves the lattice field in a
self-organized critical (SOC) state. In a critical state, among other things,
long-range correlations of suitable observables lead to power-law behavior with
respect to scaling transformations.
Self organization means that the system is driven to criticality
without fine-tuning any external parameters, i.e. solely by its intrinsic dynamics.
   
It is generally realized that financial markets, being prime examples of social systems,
exhibit SOC \cite{322494,Jensen1998-JENSCE,Bak:1997,0034-4885-62-10-201,Feigenbaum:2003}.
However, to the best of our knowledge, attempts to model those from
a microscopic point of view are rare \cite{556973}.
In \cite{1999PhyA..271..496S,RePEc:sfi:sfiwpa:500028} percolation clusters act as investors. 
Assigning random percolation probabilities, power law behavior is found for the usual
observables derived from stock market prices \cite{Ausloos:2004}.
Another example of such an attempt close to the BTW
evolution model is the work of Bartolozzi et al \cite{Bartolozzi2006b}. Though close to our
work in spirit, significant differences exist in terms of the updating strategy and
interpretation. Our implementation of our lattice field model
will produce price time series, returns and their distributions, volatility time series
and their clustering features that are all strikingly similar to historical market data.

Why would one like to have a market model in the first place? After all there are myriads of
historical data being collected every day. A good reason is the fact that all of the collected
data are merely
instances of a random draw from some probability distribution, just like one throw of a dice
gives only one number, hiding the statistics behind it.
A stochastic model on the other hand will enable us to study any
number of market instances, and collect ensembles in the language of statistical physics.
Observables, as averages endowed with errors, could be computed. Ultimately, a successful
model could provide probability distributions for future prices, and thus be an
invaluable tool for risk analysis, and the like.  

\section{\label{sec:model}Lattice model}

We consider the simplest lattice market model conceivable, a one-dimensional chain of
$n+1$ sites with labels $j=0\ldots n$, where $j$ indicates discrete time
$t=j\Delta$ in steps of some arbitrary unit $\Delta$. 
The sites are populated with a real-valued field $r$ with components $r_j\in\mathbb{R}$.
When compared to \cite{Dupoyet2010107},
we use a minimalistic topology, and also ignore the gauge field living on the
lattice links thus forgoing the minimization of arbitrage.
As it turned out it is essential to interpret the field components $r_j$ directly
as investment returns. The returns are defined as
\begin{equation}  
r_j=\log(\Phi_j/\Phi_{j-1})
\label{eq1}\end{equation}
where $\Phi_j=p_j/C$ is the price of an investment instrument,
such as a stock or index fund for example, and $C$ is a unit (currency, shares, etc).
The continuum version of (\ref{eq1}) can be surmised from taking the limit $\Delta\rightarrow 0$ in
\begin{equation}
\log({\Phi(t)}/{\Phi(t-\Delta)})=\Delta\frac{d}{dt}\log\Phi(t)+{\mathcal O}(\Delta^2)\,,
\label{eq2}\end{equation}
where $\Phi(t)=\Phi_j$ at $t=j\Delta$.

In order to endow the field $r$ with dynamics we find inspiration in the popular evolutionary
model by Bak and Sneppen \cite{PhysRevLett.71.4083,PerBak:1996,PhysRevE.53.414}.
In that context, the field
components are fitness values, say $f_j\in[0,1]$, assigned to the sites of a lattice.
The updating process consists in finding the site $j_s$ with $f_{j_s}=\min\{f_j:j=0\ldots n\}$,
i.e. the least adapted species. Then $f_{j_s}$ and the values $f_{j_s\pm 1}$ of the two
next neighbors are replaced with uniformly distributed random numbers from $[0,1]$.
This prescription, when iterated many times $s=0,1\ldots\infty$, leads to a stationary state
of the lattice field where a single perturbation can lead to a burst of activity, called an
avalanche.
The frequency distribution of avalanche sizes is found to
follow a power law. A power law is a signature feature of a critical state. Since no tuning of a
model parameter is needed, the phenomenon is known as self-organized criticality (SOC).
The model is very robust in the sense that changing the updating prescription, within
reasonable bounds, will still lead to SOC.
A rigorous discussion, containing analytical results, may be found in \cite{PhysRevE.53.414}.

\begin{figure}[ht]
\hspace{17mm}\includegraphics[angle=-90,width=101mm]{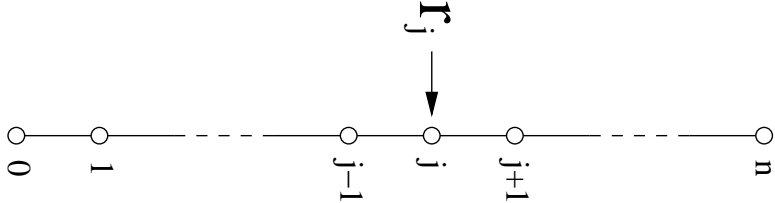}
\caption{\label{fig:lat1}Illustration of the geometry of the lattice model and the label
scheme for the sites. Periodic boundary conditions $r_{j+n+1}=r_j$ are implemented.
Updating is done on the field component with signal $V$, and its two next neighbors.}
\end{figure}

In the context of the financial market model we adopt a modified version of
Bak's updating prescription.
We select periodic boundary conditions with period $n+1$, such that
$r_{n+1}=r_{0}$ and $r_{-1}=r_{n}$.
In terms of the returns $r_j$, we define
\begin{eqnarray}
v_j&=&r_j(r_{j+1}-r_{j-1})\label{eq3a}\\
V_j&=&|v_j|\label{eq3b}
\end{eqnarray}
and call
\begin{equation}
V=\max\{V_j:j=0\ldots n\}
\label{eq4}\end{equation}
the signal of the field configuration $r$.
The updating strategy then proceeds with finding a site $j_s$
from\footnote{There is at least one, and almost always only one element in this set.}
\begin{equation}
j_s \in \{j=0\ldots n:V_j=V\}\,,
\label{eq5}\end{equation}
and then replacing the returns on sites $j_s$ and its two neighbors according to
\begin{equation}
r_{j_s} \leftarrow x_0\quad\mbox{and}\quad r_{j_s\pm 1} \leftarrow x_{\pm 1}
\label{eq6}\end{equation}
where $x_0$ and $x_{\pm 1}$ are three random numbers drawn from a normal
distribution $p(x)\propto\exp(-x^2/2w)$
while enforcing the constraint $x_{-1}+x_0+x_{+1}=0$.
The variance $w$ is a parameter.
Figure~\ref{fig:lat1} illustrates the situation.

Within reason, we have experimented with numerous alternative definitions for the signal,
defined through (\ref{eq3a},\ref{eq3b},\ref{eq4}).
Overall, it appears that Bak updating is very robust and SOC is easily achieved.
However, the choice (\ref{eq3a},\ref{eq3b},\ref{eq4}) proved to best match the stylized features of
historical financial market data.

Although (\ref{eq3a},\ref{eq3b},\ref{eq4}) were mostly selected on empirical grounds, in
retrospect, more motivation may be provided: Note that (\ref{eq3a}) is just a
discretized version of
\begin{equation}
v(t)=r(t)2\Delta\frac{dr(t)}{dt}=\Delta\frac{d}{dt}r(t)^2
\label{eq7}\end{equation}
where $r(t)$, in view of (\ref{eq1}) and (\ref{eq2}), has been identified with
\begin{equation}
r(t)=\Delta\frac{d}{dt}\log\Phi(t)\,.
\label{eq8}\end{equation}
In finance, an established approach is to treat the returns $r(t)$ as a stochastic
process \cite{Engle:1982:ARCH}. Returns are typically modeled by a (generalized)
Wiener process, i.e. assuming normal distributed random variables with a time dependent
(random) variance
\begin{equation}
W(t)=E[r(t)^2] - E[r(t)]^2\,,
\label{eq9}\end{equation}
where $E[\cdots]$ indicates the stochastic expected value. 
Discretized versions of the time derivatives of the two terms are
\begin{eqnarray}
\frac{d}{dt} E[r(t)^2] = 2E[r(t)r^\prime(t)]
& \simeq & \langle r_j(r_{j+1}-r_{j-1})\rangle \Delta^{-1} \label{eq10a}\\
\frac{d}{dt}E[r(t)]^2 = 2E[r(t)]E[r^\prime(t)]
& \simeq & \langle r_j \rangle\langle r_{j+1}-r_{j-1}\rangle \Delta^{-1} \label{eq10b}\,.
\end{eqnarray}
On the right-hand sides, we have changed the notation for the expectation value from
$E[\cdots]$ to $\langle \cdots \rangle$, the latter indicating the averages for
lattice generated returns.
In the current simulation, standard stochastic financial modeling dictates
that $\langle r_j \rangle$ be independent of time. This implies that the discretized
part of equation (12) is exactly equal to zero.

Therefore, the dynamics of the lattice model is driven by
eliminating extreme, sudden, changes of the variance on the returns deemed
`unfit' in the spirit of \cite{PhysRevLett.71.4083}.
Combining (\ref{eq9}) and (\ref{eq10a},\ref{eq10b}) the lattice version of the latter
turns out to be
\begin{equation}
\frac{dW(t)}{dt} \simeq [\cov(r_{j},r_{j+1})-\cov(r_{j},r_{j-1})]\Delta^{-1} \,,
\end{equation}
where $\cov(r_{\alpha},r_{\beta})=
\langle r_{\alpha}r_{\beta}\rangle-\langle r_{\alpha}\rangle\langle r_{\beta}\rangle$
is the covariance of the two random variables.
Hence, extreme changes of the covariance of returns between adjacent time slices are
discouraged as part of the dynamics of the market model.

As a side remark we comment on the use of the two-step time derivative to approximate
$r^\prime(t)$ in (\ref{eq10a}) and (\ref{eq10b}) as opposed to employing one-step
forward $(r_{j+1}-r_j)\Delta^{-1}$, or backward $(r_j-r_{j-1})\Delta^{-1}$,
discretizations. In both cases the interaction (\ref{eq3a}), which drives the lattice dynamics,
would mutate from a proper next-nearest-neighbor coupling to a term
dominated by self-interactions $\propto r_j^2$. This does not lead to a sensible physical
system, and demonstrably gives absurd results in a numerical simulation.
The corresponding approximations to the time derivative of the variance $dW/dt$ are
$\simeq [\cov(r_{j+1},r_j)-\var(r_j)]2\Delta^{-1}$ and
$\simeq [\var(r_j)-\cov(r_j,r_{j-1})]2\Delta^{-1}$, respectively. 
Those are variance driven and, equally, not suitable to define market dynamics.

Finally, we should caution that the purpose of the above exposition is only to provide some
motivation. In fact we should expect that the returns time series has fractal geometry.
This makes perfect sense on a discretized time lattice, with the usual implications
\cite{Mandelbrot:1997,Sornette:2000}, but it certainly discourages using the concept of
continuous time derivatives on everyday manifestations of market data.
In this vein the lattice model, in combination with
next-nearest-neighbor interactions, appears to be a rational approach.
 
\section{\label{sec:simulation}Simulation}

\begin{figure}[ht]
\hspace{17mm}\includegraphics[angle=90,width=101mm]{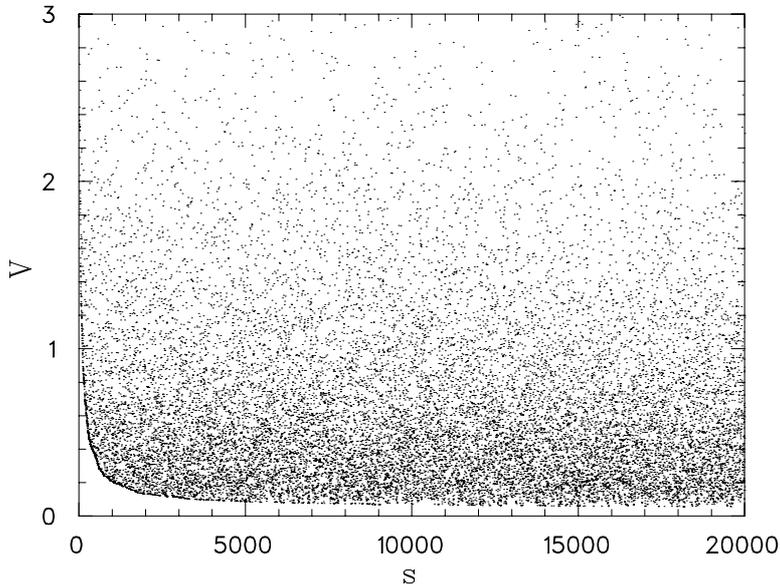}
\caption{\label{fig:btw-Vs}Updating evolution of the signal $V$ versus the simulation time $s$.
The signal at $s=0$ is due to a random start of the lattice field configuration.
(Its value is $=4.74$.)}
\end{figure}

For good measure, we have chosen a lattice with $n=780$ time steps.
Starting with a random lattice field $r$
we show in Fig.~\ref{fig:btw-Vs} an example of the updating evolution of the signal
$V=\max\{|r_j(r_{j+1}-r_{j-1})|:j=0\ldots n\}$ versus the simulation `time' $s$.
Clearly visible is a significant drop of the lower envelope of $V(s)$ as $s$
approaches $\approx 10000$. Beyond that the distribution of signal dots appears to
have stabilized at around $\approx 20000$, at least this is the visual impression.

\begin{figure}[p]
\hspace{17mm}\includegraphics[angle=90,width=101mm]{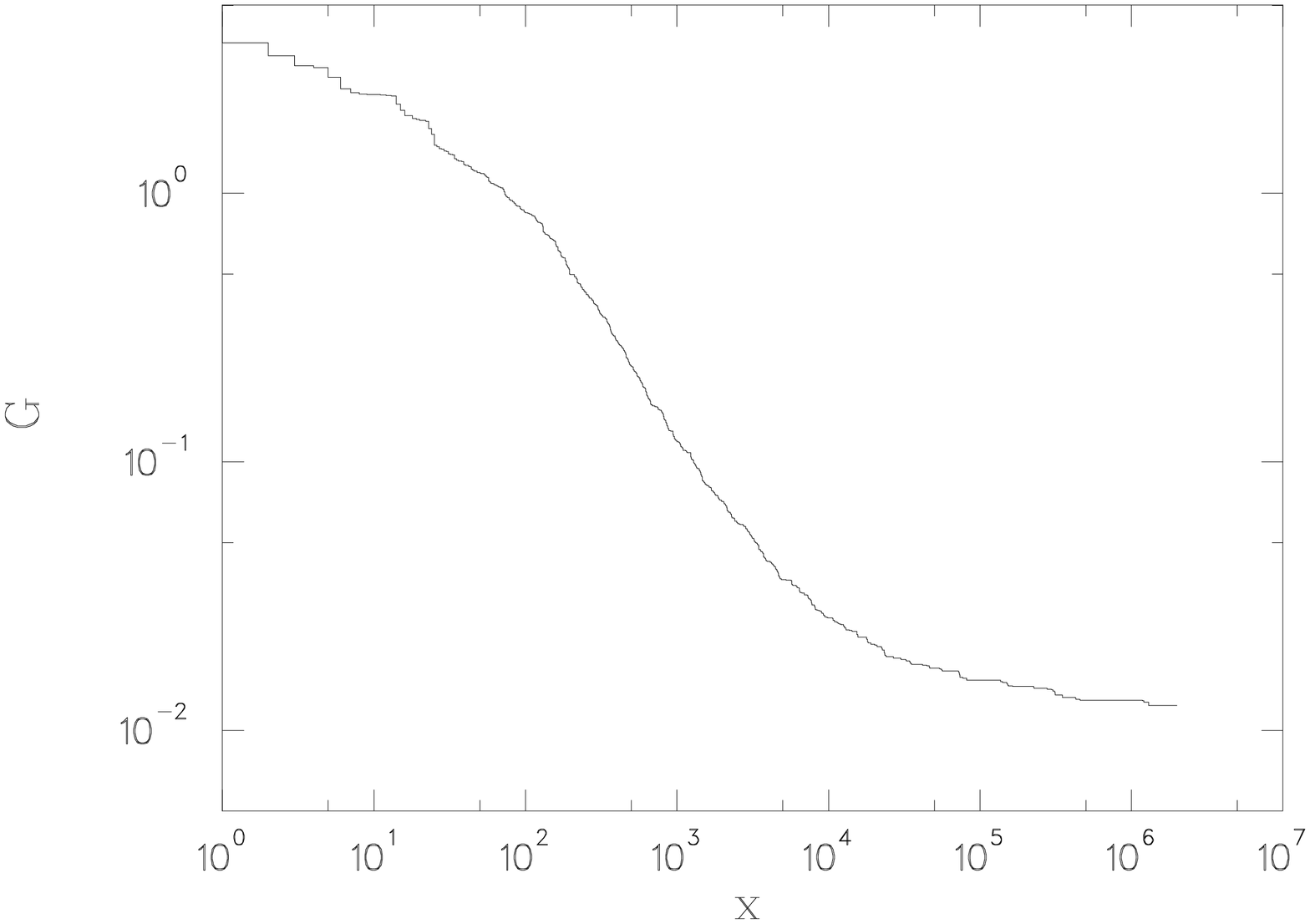}
\caption{\label{fig:btw-gp}Plot of the gap function (\protect\ref{eq11})
of Fig.~\protect\ref{fig:btw-Vs}, but up to $x=2\times 10^6$.}
\end{figure}

Following \cite{PhysRevE.53.414} we take a closer look at the
envelope and define the `gap' function 
\begin{equation}
G(x)=\min\{V(s):s\in{\mathbb N}\cup\{0\} \;\mbox{and}\; s\le x \}
\quad\mbox{with}\quad x\in{\mathbb R}^{+}\cup\{0\}\,,
\label{eq11}\end{equation}
which is meant to trace the lower envelope of the signal.
By construction, $G(x)$ is a decreasing piecewise constant
function with discontinuities at certain discrete
values $x_k, k\in{\mathbb N}$. Adding $x_0=0$ and assuming an ordered sequence
the length of a plateau is $\Lambda_k=x_{k}-x_{k-1}$, and its height is $G(x_{k-1})$. 
By definition, we say that an avalanche of length $\Lambda_k$ starts at $x_{k-1}$ and
ends at $x_k$. At $s=x_{k-1}$ all lattice sites have local signals $V_j\le G(x_{k-1})$,
see (\ref{eq4}).
As long as the avalanche lasts, there is at least one lattice site with a local signal
larger than $G(x_{k-1})$ and thus the updating activity continues until $s=x_{k}$. 
Since the gap function is decreasing and bounded from below by zero it will eventually
approach a constant $\lim_{x\rightarrow\infty}G(x)=G_C$. In this regime the avalanche size
diverges and the system has reached the desired state of criticality \cite{PhysRevE.53.414}.
An example of $G(x)$ from our simulation is shown in Fig.~\ref{fig:btw-gp}.
It corresponds to the data of Fig.~\ref{fig:btw-Vs} but up to much larger simulation times.
The evidence points to a critical value $G_C$ with $0\le G_C \lesssim 0.01$.
Note that $G_C=0$ is a possibility, we do not know if it is realized.
Unlike for the evolution model \cite{PhysRevE.53.414} no analytic results are available.
Strictly speaking, the above narrative applies to the thermodynamic limit, implying
a lattice with infinitely many sites. Thus, in principle, we expect finite-size
effects to afflict our simulation. However, because subsequent results are very sensible
we don't expect those to be an obstacle to practical application of this model.
\begin{figure}[p]
\hspace{17mm}\includegraphics[angle=90,width=101mm]{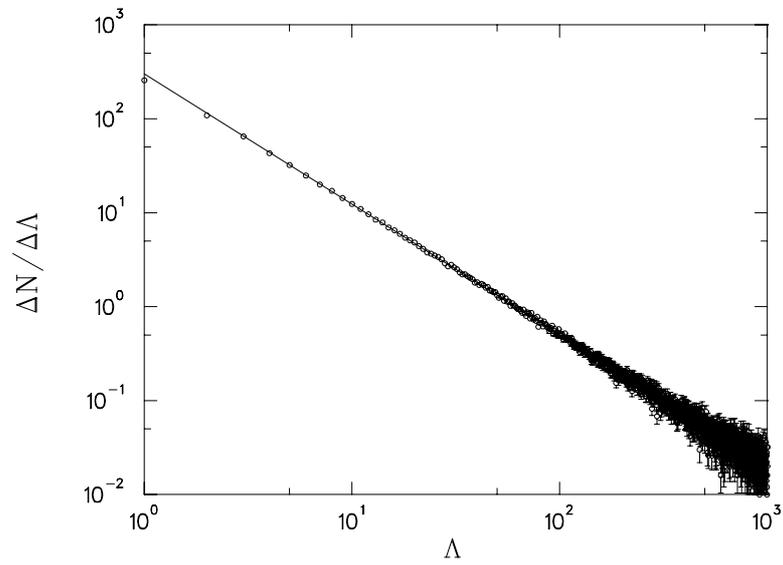}
\caption{\label{fig:btw-ap}Frequency distribution of avalanche sizes. The straight
line corresponds to a power law fit.}
\end{figure}

A signature feature of a critical system is scale invariance, implying
power law behavior of certain quantities.
Again following \cite{PhysRevE.53.414}, we display in Fig.~\ref{fig:btw-ap}
the frequency distribution of the avalanche sizes $\Delta N/\Delta\Lambda$
where $\Delta\Lambda$ is a binning interval for the avalanche sizes and
$\Delta N$ is the count of avalanches within that interval. 
We have used 10000 bins with a binning interval of $\Delta\Lambda=1$.
The data points come from an ensemble average over $2000$ independent lattice simulations
with $2\times 10^6$ update steps each.
This allows one to calculate statistical errors, also displayed in Fig.~\ref{fig:btw-ap}.
A power law behavior is beyond doubt. A least-$\chi^2$ fit including data points in
the interval $10^1 \leq \Lambda \leq 10^2$ gives $\Delta N/\Delta\Lambda = 301\, \Lambda^{-1.39}$.
Integrating this gives a total number of $N\approx 780$ avalanches out of
a run with $2\times 10^6$ updates. 
However, the integrated average $N$ is clearly less than the actual avalanche count per simulation
due to the fact that the averaging operation gives rise to counts less than unity,
which is realistically unfeasible. An analysis over many simulations gives a more accurate count
average where $780 \le N \le 1000$ for a single run
with 2-million updates. Hence long avalanches clearly are key to the updating process.

\begin{figure}[p]
\hspace{17mm}\includegraphics[angle=90,width=101mm]{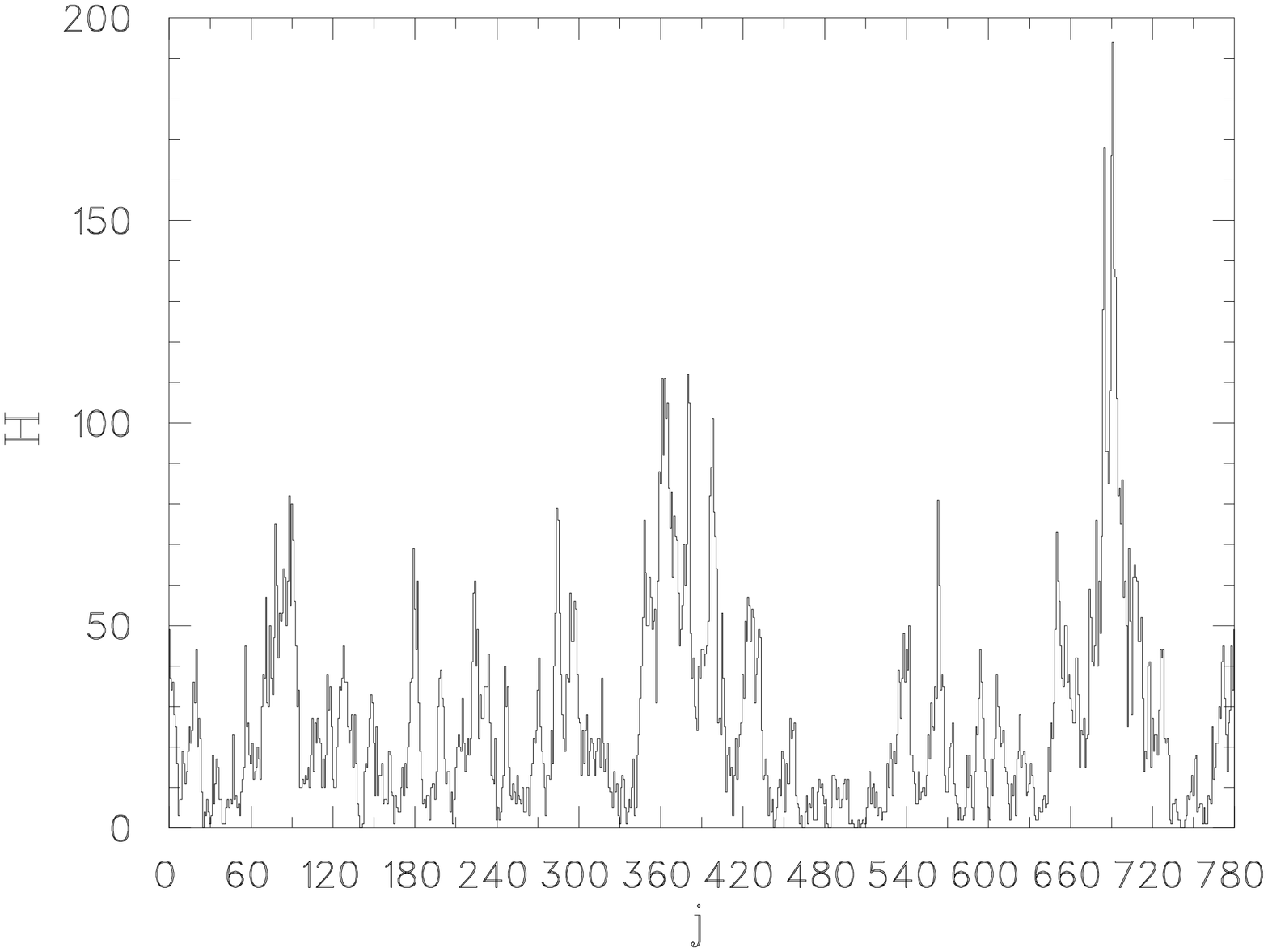}
\caption{\label{fig:btw-ac}Plot of the activity across the lattice sites after
20000 update steps. The appearance of peaks signals the emergence of avalanche dynamics.}
\end{figure}

In Fig.~\ref{fig:btw-ac} we display the activity of the lattice sites during an
updating sequence.
By definition, $H$ is the number of times a lattice site $j$ has
been visited by an updating hit since the start of the simulation.
Initially, with random $r_j$ assigned to the sites
the signal $V$ tends to visit each site with comparable probability, leading to a
flat activity plot. The plot in Fig.~\ref{fig:btw-ac} reflects the activity
after 20000 updates. A glance at Fig.~\ref{fig:btw-Vs}, and Fig.~\ref{fig:btw-gp},
reveals that this corresponds to the onset of criticality. The peak-like
structures in the activity plot give evidence of the emergence of avalanches with
larger sizes.   
\begin{figure}[p]
\hspace{17mm}\includegraphics[angle=90,width=101mm]{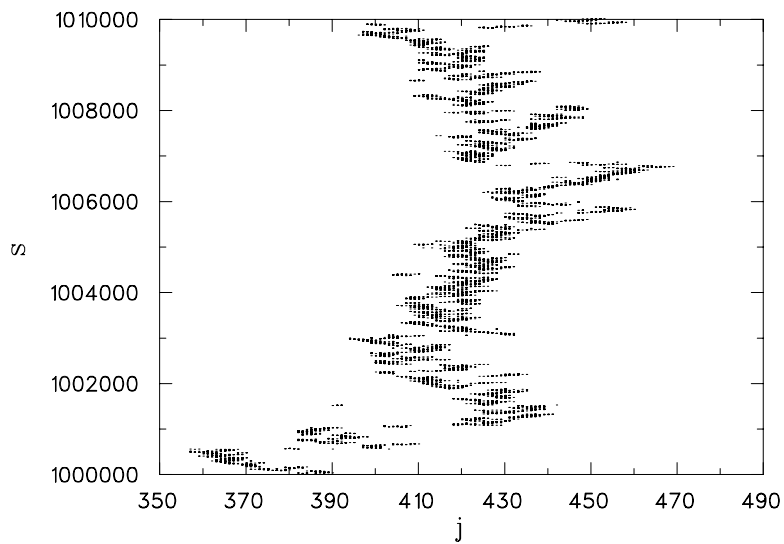}
\caption{\label{fig:btw-az}A zoom window on the updating evolution. Each plot symbol (-)
at site $j$ and simulation time $s$ indicates an updating hit.}
\end{figure}

A closely related plot shown in Fig.~\ref{fig:btw-az} shows a zoom window on the
simulation time dependence of the updated sites around $j\approx 420$
and $s\approx 10^6$.
Each plot symbol in Fig.~\ref{fig:btw-az} indicates an updating event of site $j$ at simulation
time $s$. Disconnected sets of dots clearly show the presence of avalanches.
By visual inspection the curve is fractal in nature, a signature feature of
complexity. In the context of the evolution model \cite{PhysRevLett.71.4083}
the term `punctuated equilibrium' has been used to describe a similar observation.

\begin{figure}[h!]
\hspace{17mm}\includegraphics[angle=90,width=101mm]{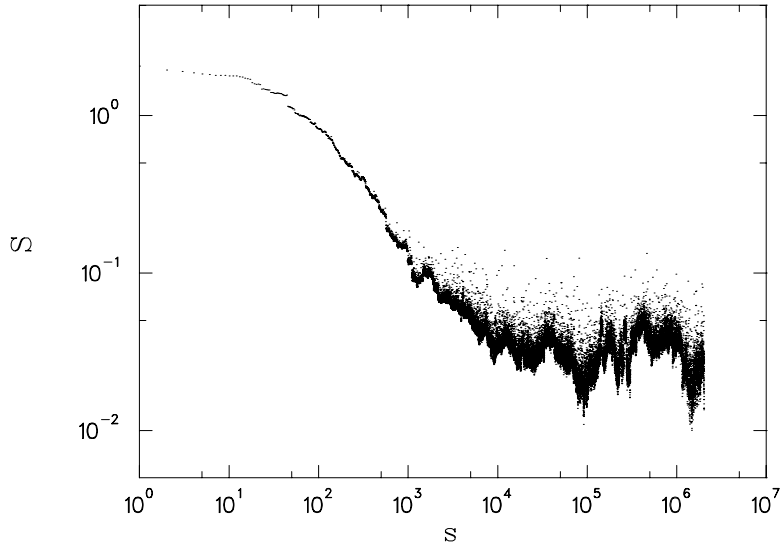}
\caption{\label{fig:btw-en}Entropy $S$ of the exponentiated returns $R_j=\exp(r_j)$
as a function of the simulation time $s$.}
\end{figure}

Finally, we have used an entropy-like quantity to monitor the approach of the field
towards a complex state. Using the exponentiated returns $R_j=\exp(r_j)$ define
\begin{equation}
S=\frac{1}{n}\sum_{j=1}^{n}R_j\log R_j\,.
\label{eq21}\end{equation}
Then Fig.~\ref{fig:btw-en}, in which is displayed $S$ versus the simulation time $s$,  
shows that the initially random system becomes organized, again at around $s=20000$.
From then on the information content of the field has stabilized, as indicated by
bounded fluctuations of $S$ between $10^{-1}$ and $10^{-2}$.

\section{\label{sec:results}Results}

As mentioned above, with our lattice geometry and size, it takes at least 20000 updating
hits to reach a critical state. The lattice has $n=780$ time intervals, and we use $w=1$.
To obtain the results discussed in this section we have used $4\times 10^6$ initial
updates before collecting field configurations from independent simulations.

\subsection{Gains distribution}
 
\begin{figure}[p]
\center
\includegraphics[angle=90,height=84mm]{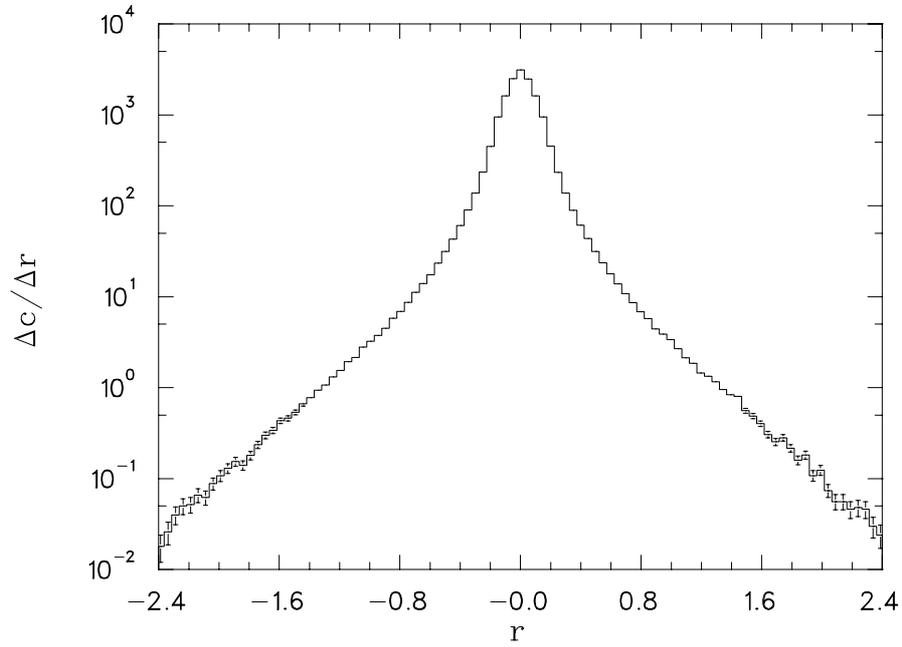}
\caption{\label{fig:rd}Lattice returns (gains) distribution. $\Delta c/\Delta r$ is the
number of returns of size $r$ at a binning interval of $\Delta r = 0.05$.}
\end{figure}
\begin{figure}[p]
\center
\includegraphics[angle=90,height=84mm]{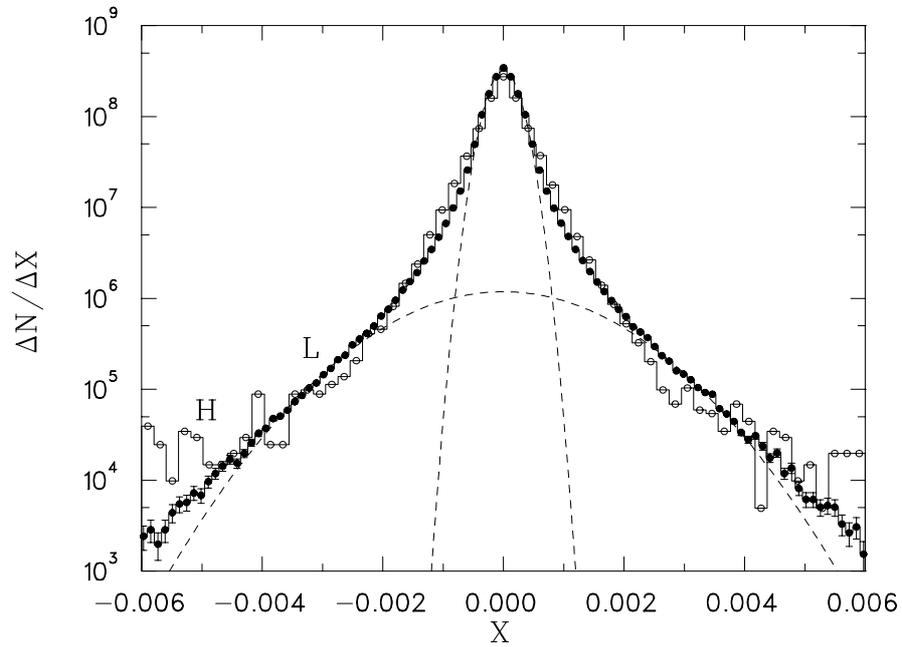}
\caption{\label{fig:hd}Historical NASDAQ returns (gains) distribution (open circles, H)
compared to the results of the lattice simulation (filled circles, L). The dashed lines
are Gaussian fits to the center and the tails, respectively, of the lattice data.}
\end{figure}

The advantage of working with a stochastic model is that observables can be
estimated from ensembles, i.e. multiple realizations of market time series.
Thus we can take a look at the gains distribution, defined as the frequency
plot of returns against the size of the returns. The lattice data displayed
in Fig.~\ref{fig:rd} are averages from $10000$ simulations,
taking one sample after $4\times 10^6$ updates each.
The binning interval for the
returns is $\Delta r=0.05$ and the number of counts per interval $\Delta c/\Delta r$
is normalized such that the total number of counts is $n=780$.
Statistical errors are visible at the tails of the distribution.
A well-known feature of gains distributions is that they exhibit `fat tails', meaning
that for extreme values of $r$ the distribution is considerably enhanced over a normal (Gaussian)
distribution. In \cite{Dupoyet2010107}, where we studied the effects of arbitrage using
a local gauge field, this feature was only obtained after analyzing the returns
performing a weighted time average of past returns. In the present model the fat tails
distribution emerges naturally. Since the returns field develops into a critical state the
resulting scale invariance implies loss of memory of past holding patterns.

The same gains distribution is displayed in Fig.~\ref{fig:hd} along with historical
market data from the NASDAQ index compiled from minute data between
2005-Aug-26 and 2008-Aug-25 \cite{Finam:2008}.
Scale factors as in $\Delta X=2.4\times 10^{-3}\Delta r$
and $\Delta N/ \Delta X= 1.1\times 10^5 \Delta c/ \Delta r$ have been applied to
match the historical data. The dashed lines correspond to Gaussian distributions
fit to the lattice data in the center ($7$ points) and the tails ($2\times 38$ points),
respectively. The Gaussian distribution is clearly ruled out. We observe that the lattice model accounts
remarkably well to the empirical gains distribution over many orders of magnitude. 

\subsection{Time series}

\begin{figure}[p]
\includegraphics[angle=90,width=66mm]{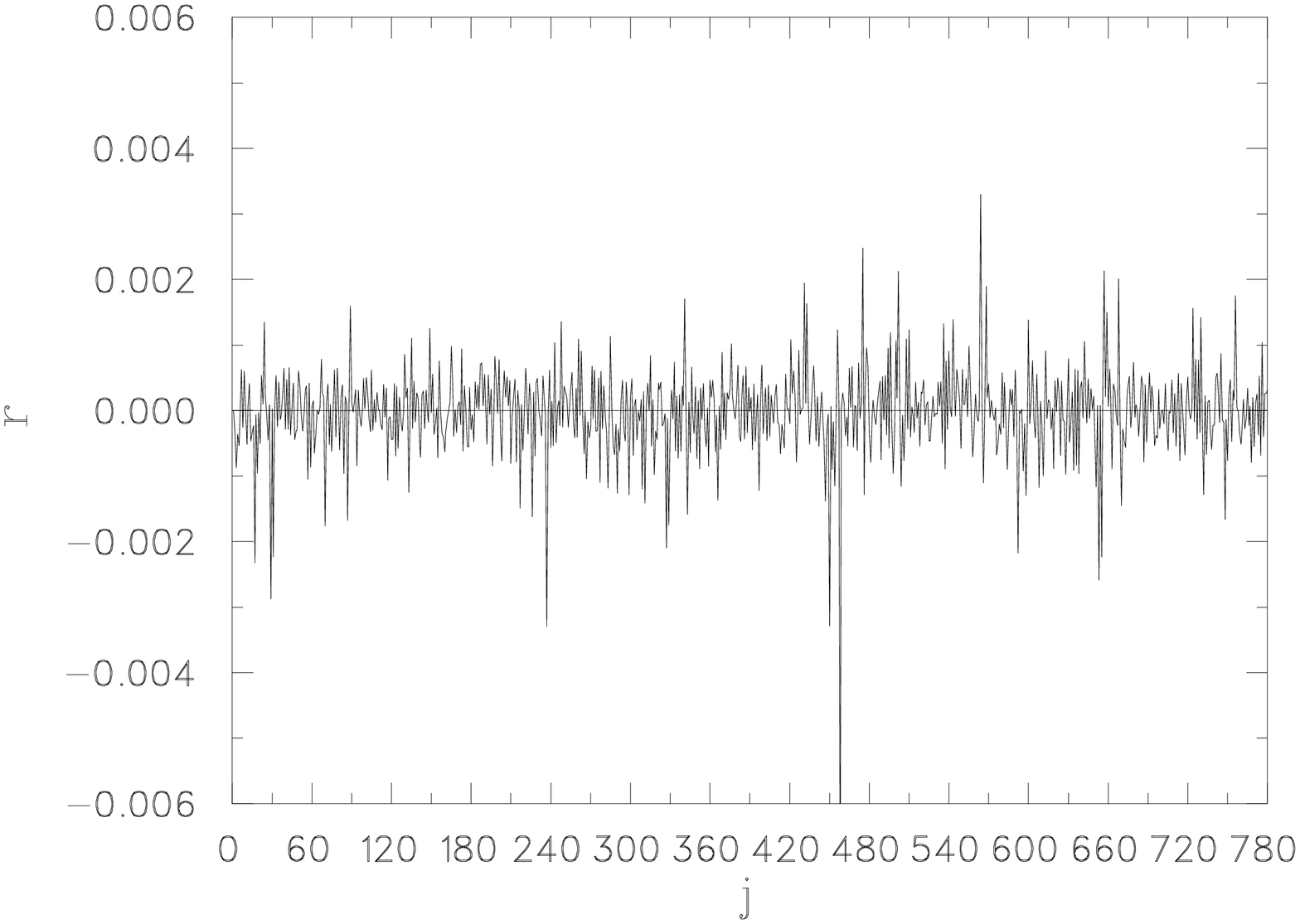}\hspace{3mm}
\includegraphics[angle=90,width=66mm]{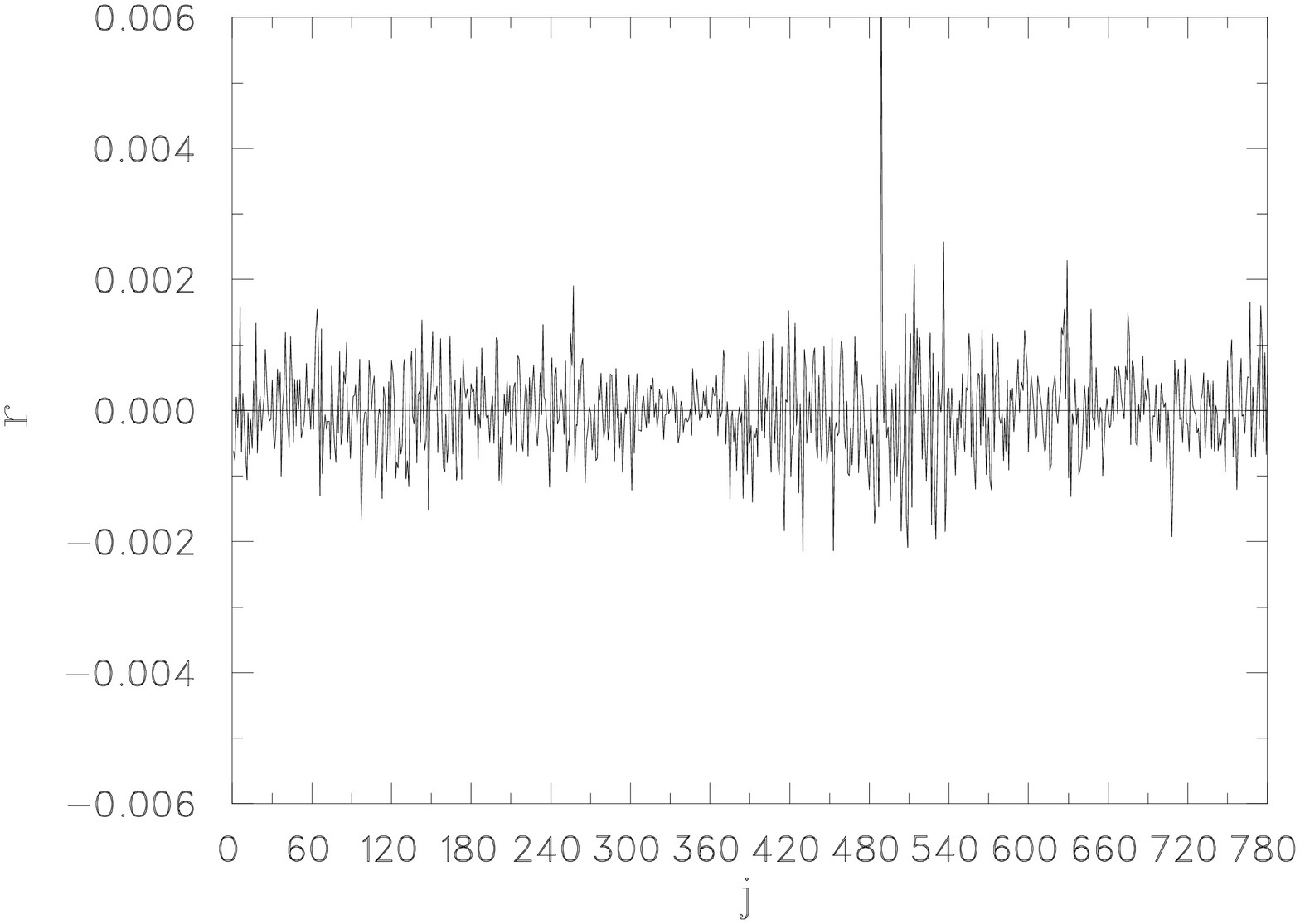}\vspace{4mm}\\
\rule{2mm}{0mm}
\includegraphics[angle=90,width=63mm]{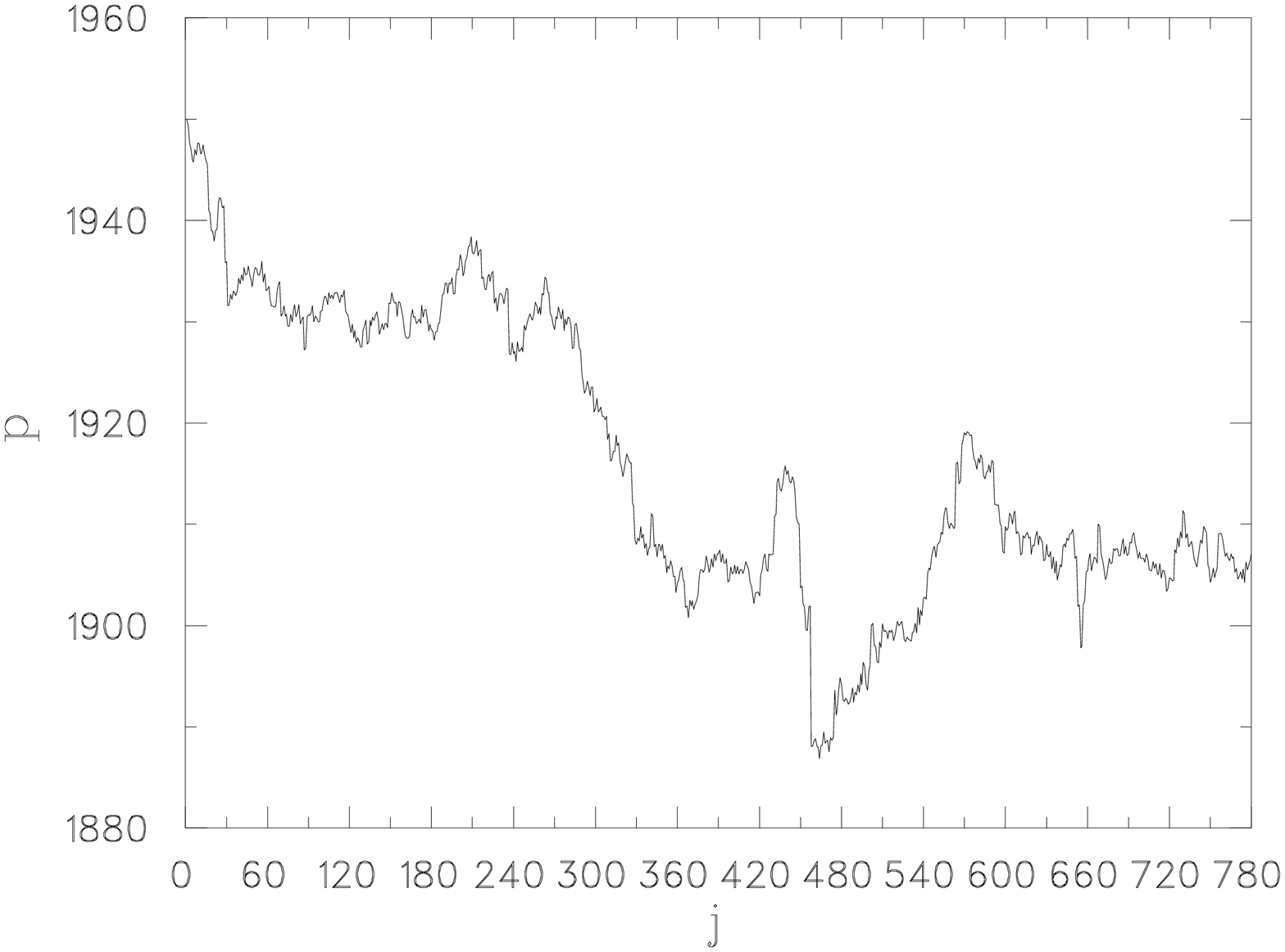}\hspace{6mm}
\includegraphics[angle=90,width=63mm]{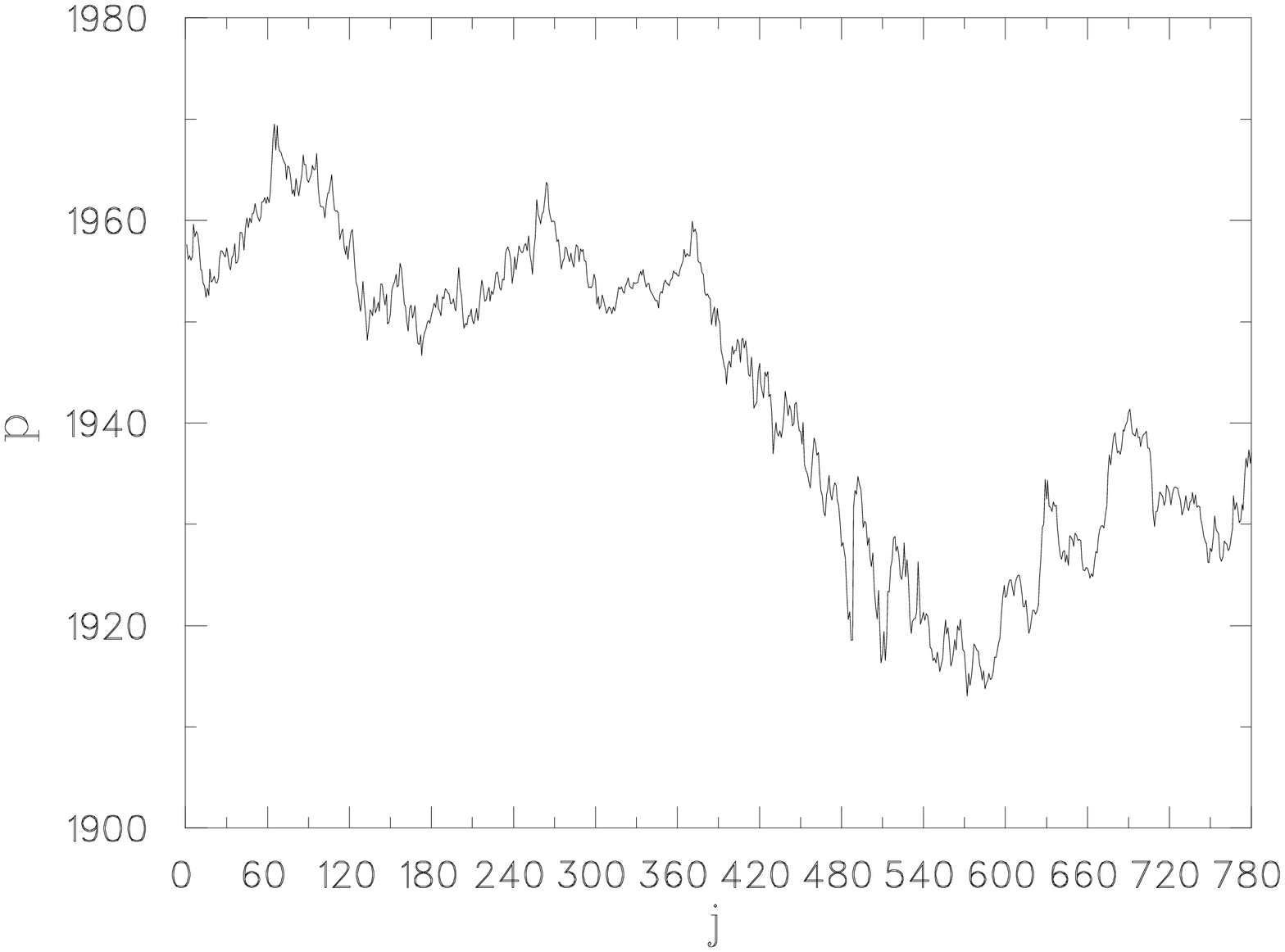}\vspace{4mm}\\
\rule{3mm}{0mm}
\includegraphics[angle=90,width=62mm]{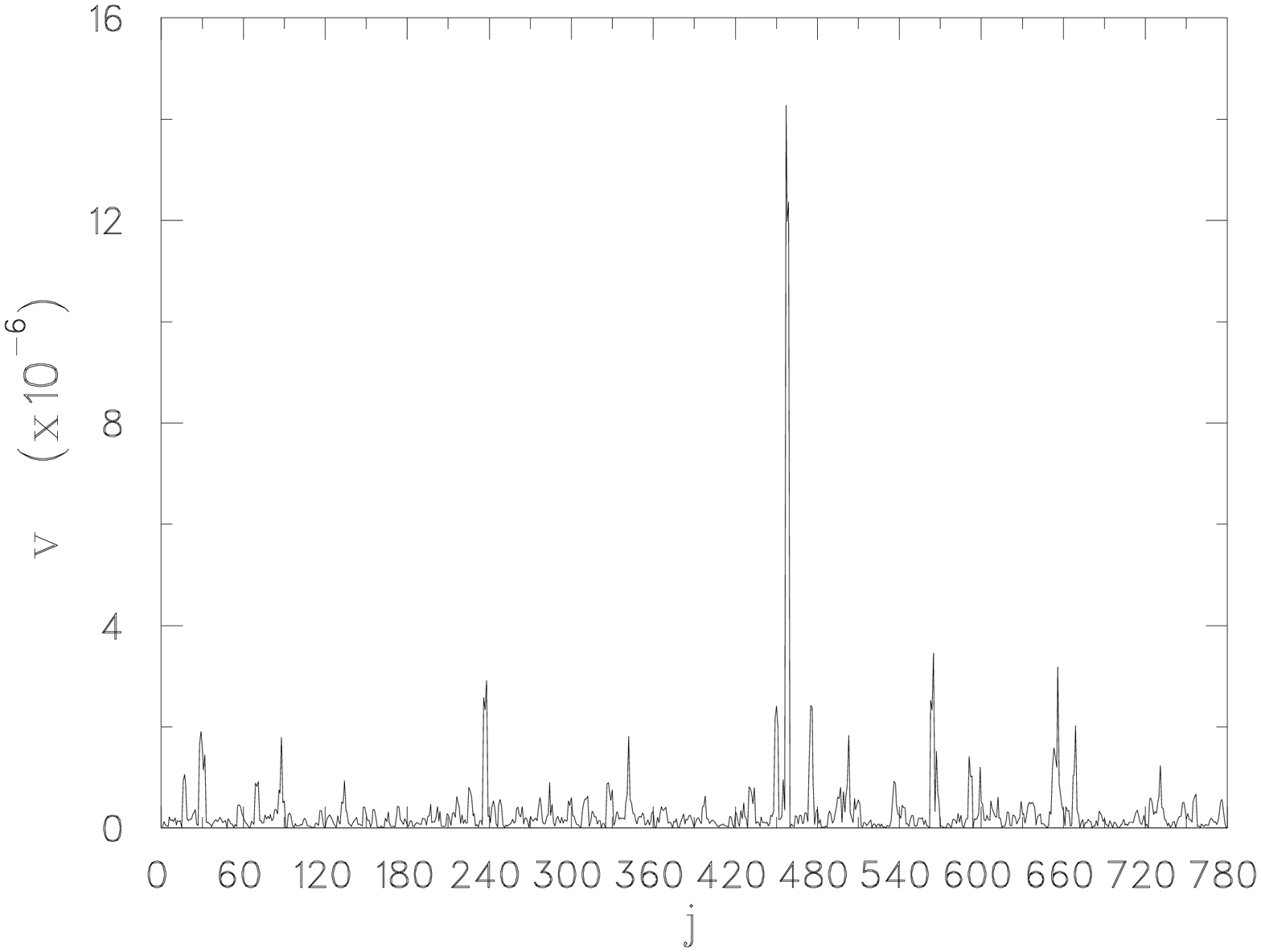}\hspace{7mm}
\includegraphics[angle=90,width=62mm]{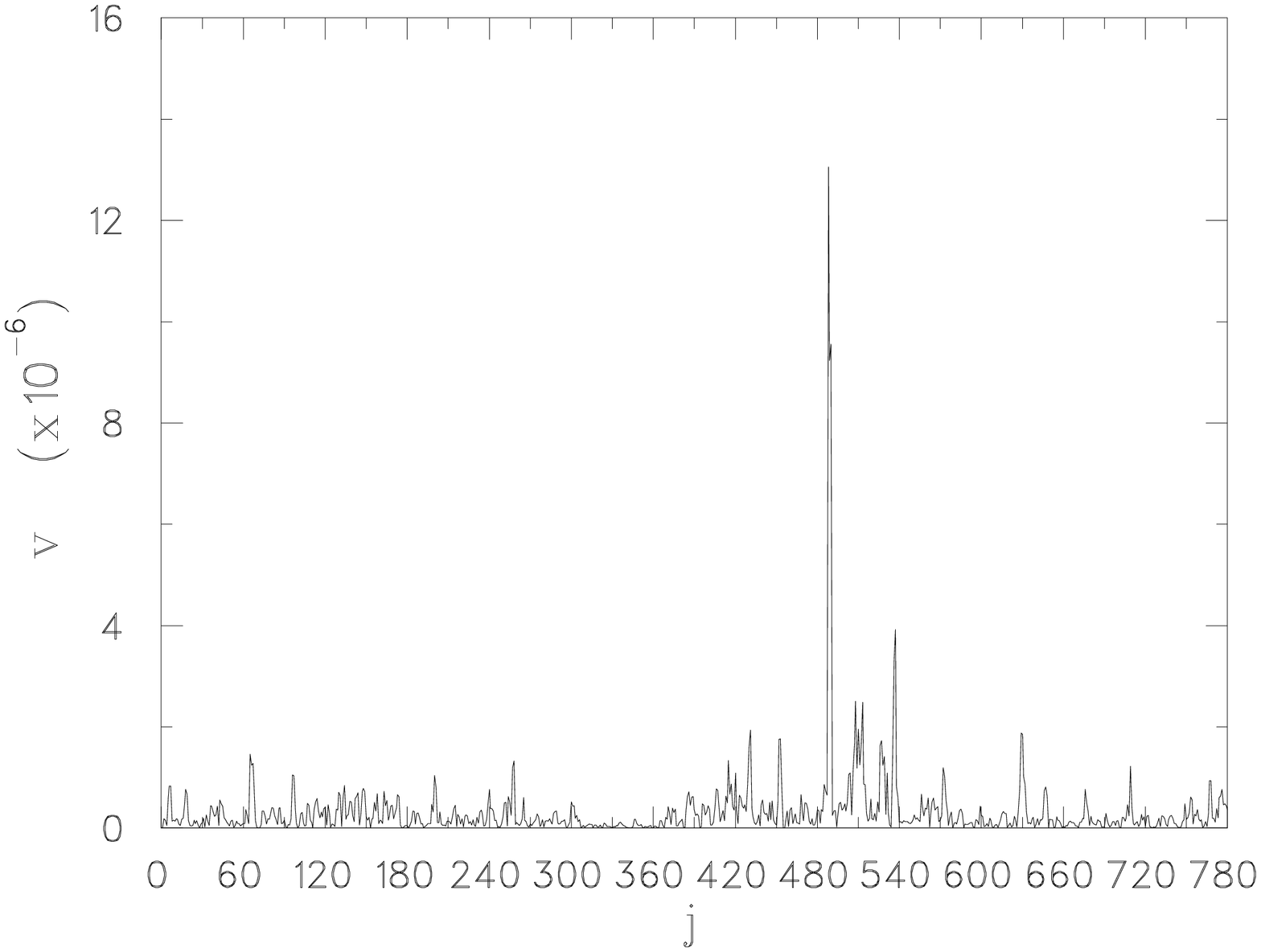}
\caption{\label{fig:flh1}Set 1: Lattice generated times series of returns $r$, prices $p$,
and volatility $v$ are shown in the left column. The three panels on the right show
selected historical $r,p,v$ times series from the NASDAQ index.}
\end{figure}

\begin{figure}[p]
\includegraphics[angle=90,width=66mm]{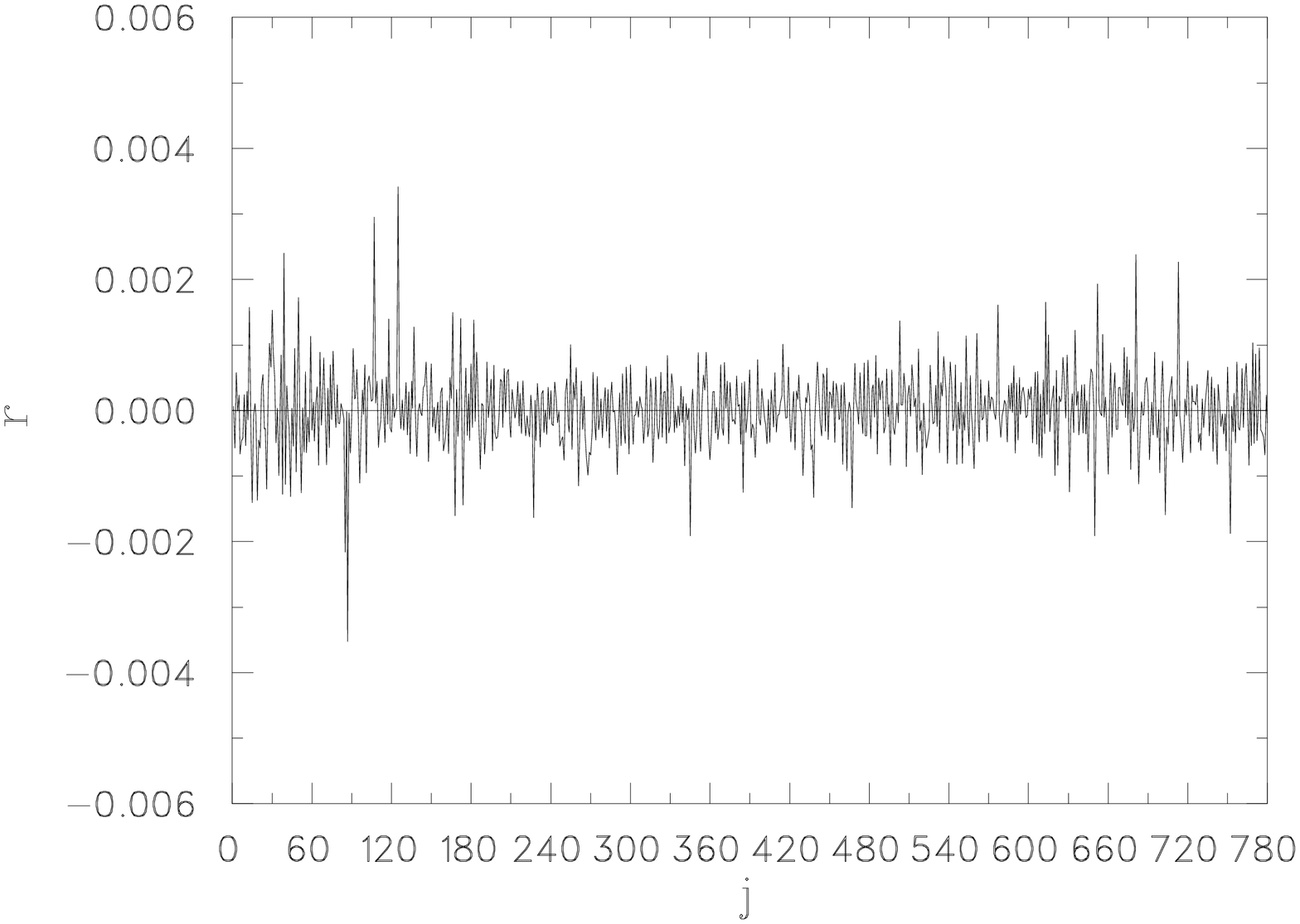}\hspace{3mm}
\includegraphics[angle=90,width=66mm]{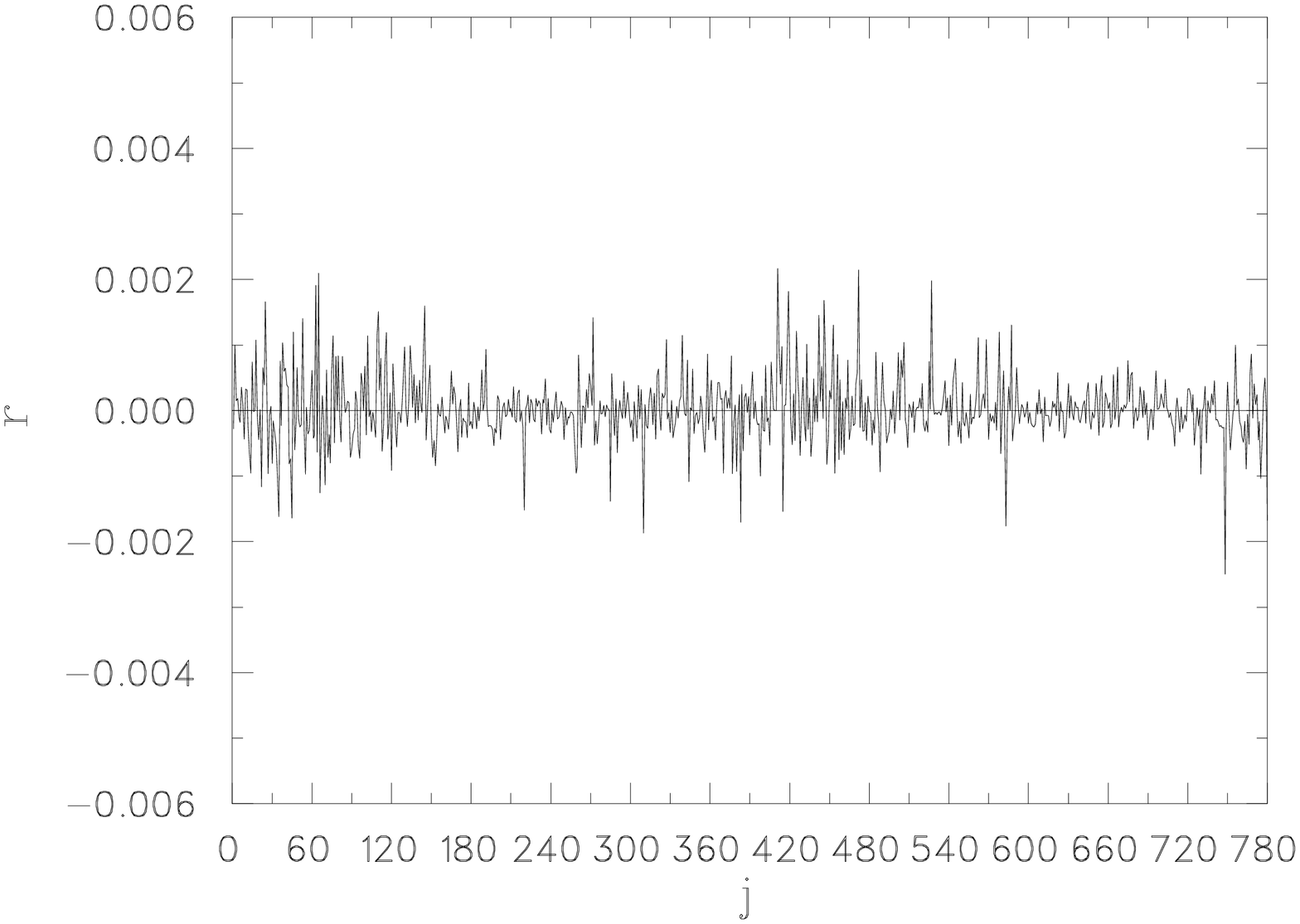}\vspace{4mm}\\
\rule{2mm}{0mm}
\includegraphics[angle=90,width=63mm]{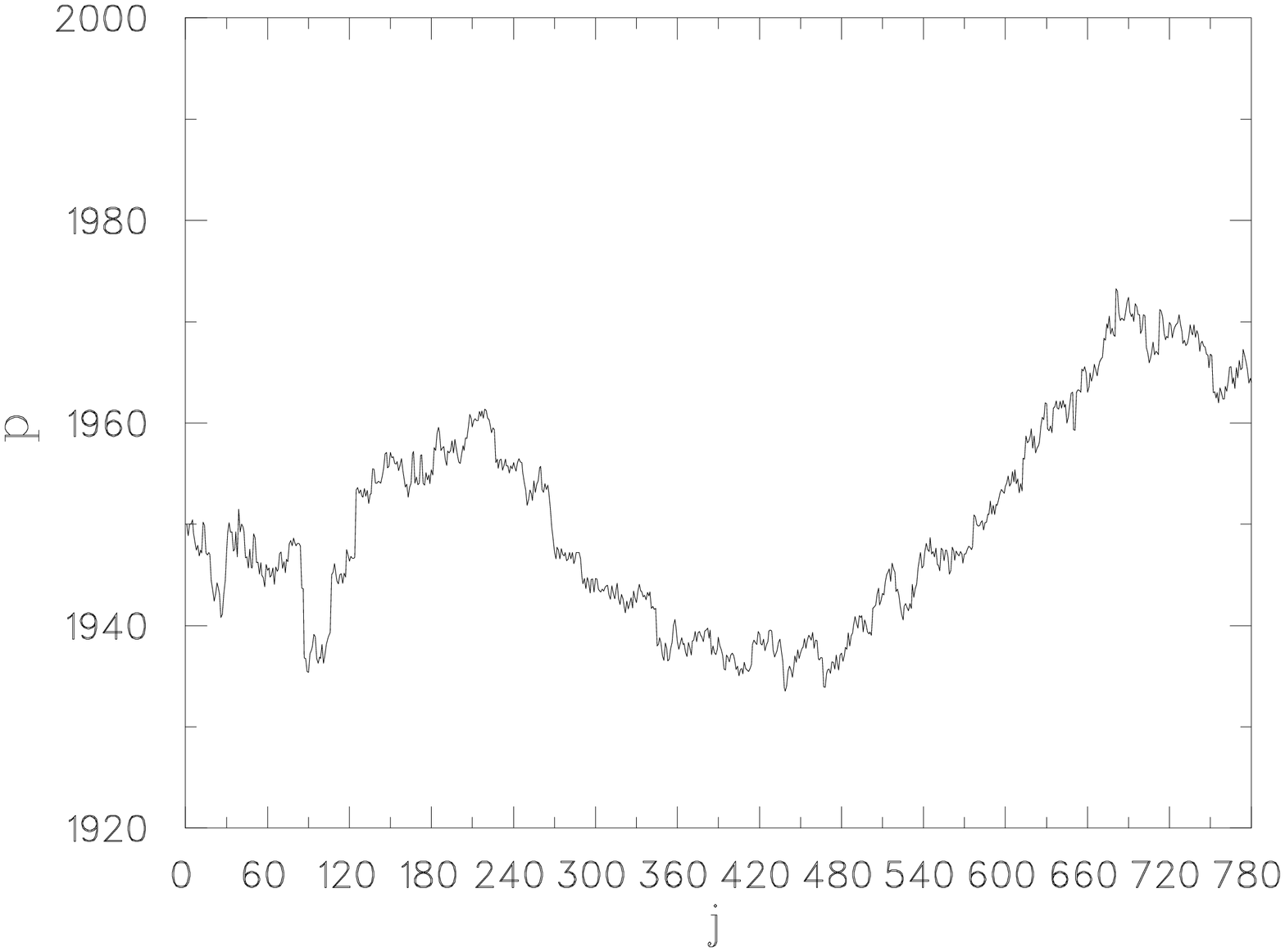}\hspace{6mm}
\includegraphics[angle=90,width=63mm]{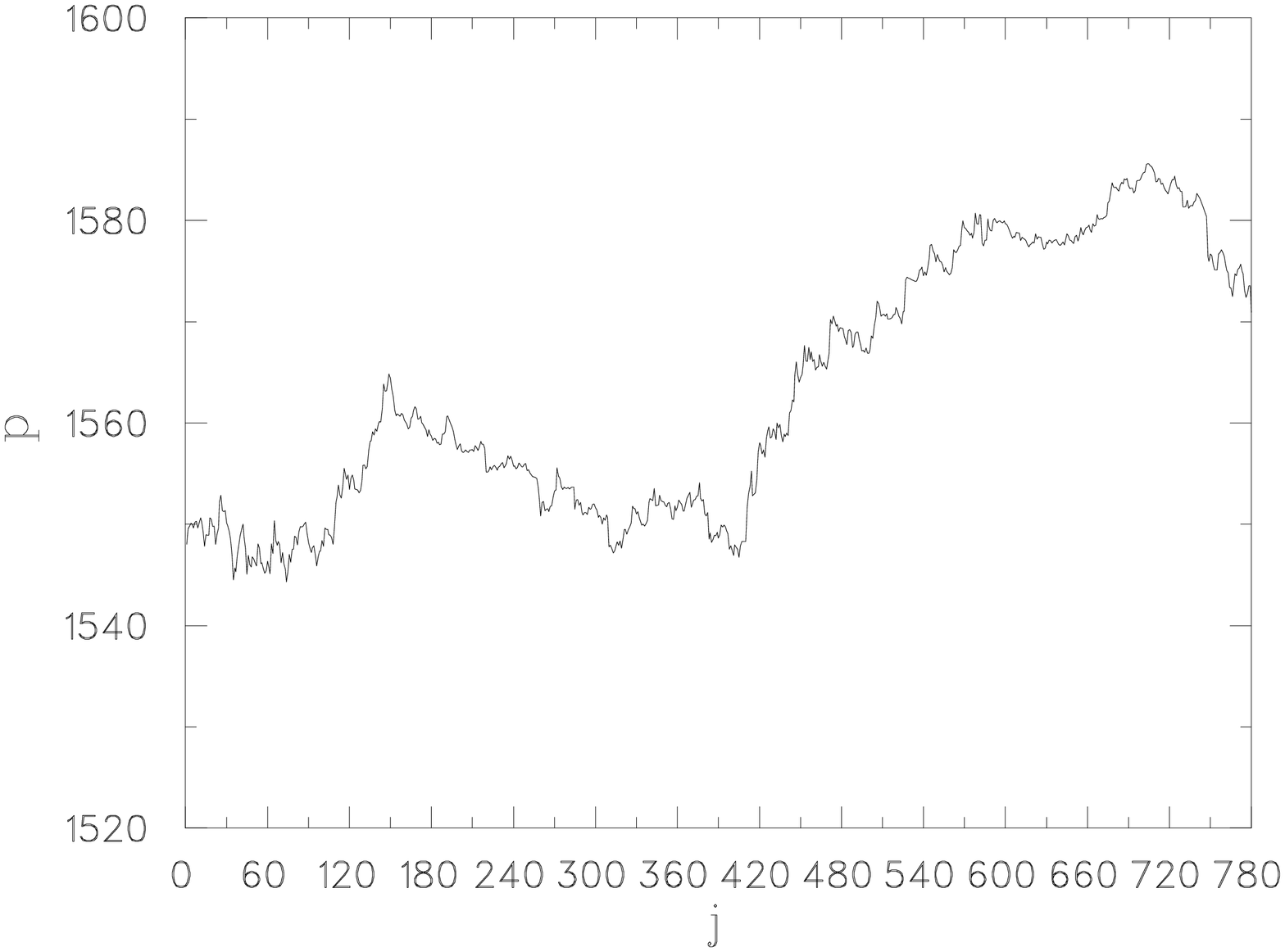}\vspace{4mm}\\
\rule{3mm}{0mm}
\includegraphics[angle=90,width=62mm]{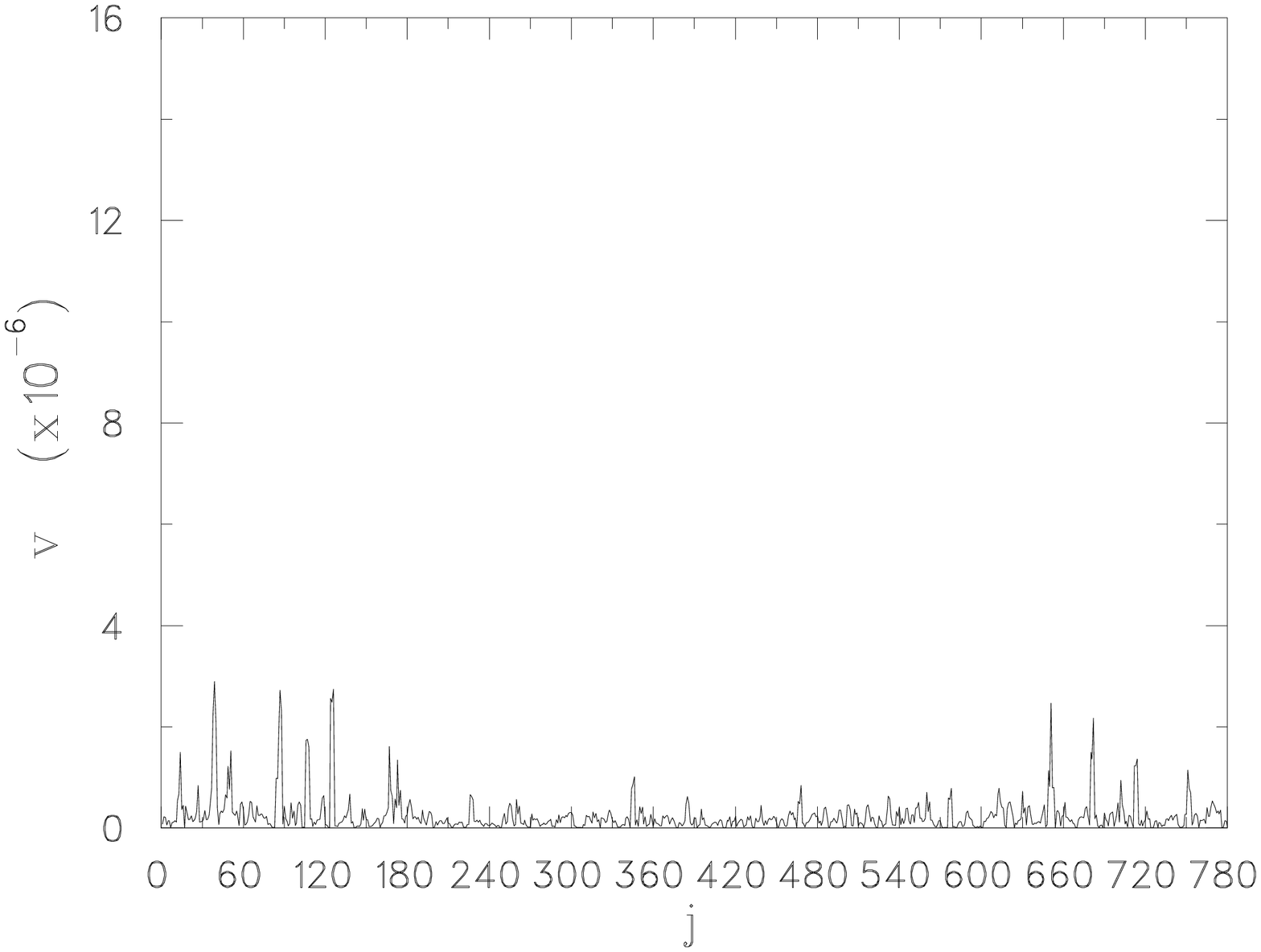}\hspace{7mm}
\includegraphics[angle=90,width=62mm]{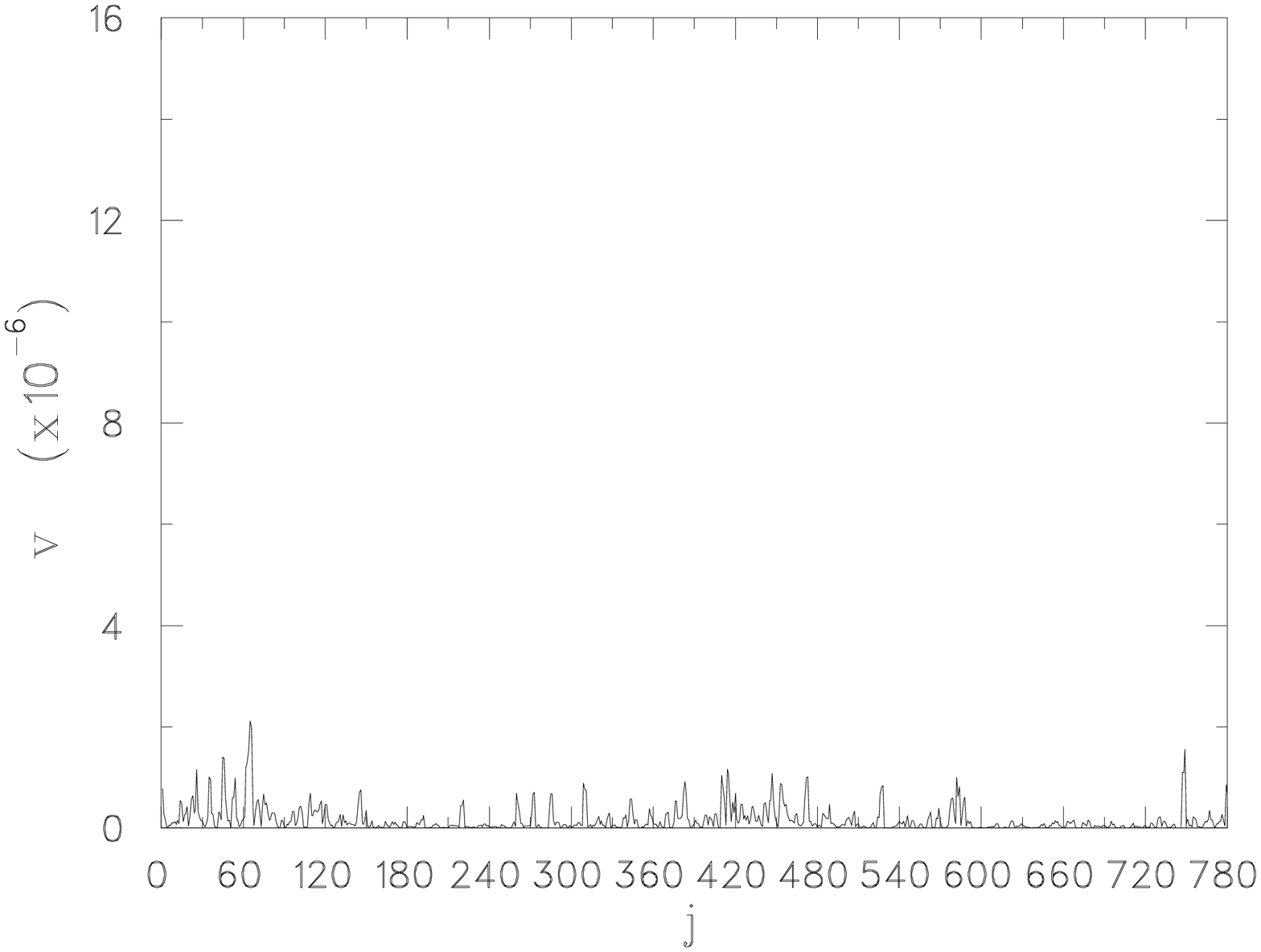}
\caption{\label{fig:flh2}Set 2: See caption of Fig.~\protect\ref{fig:flh1}.}
\end{figure}

\begin{figure}[p]
\includegraphics[angle=90,width=66mm]{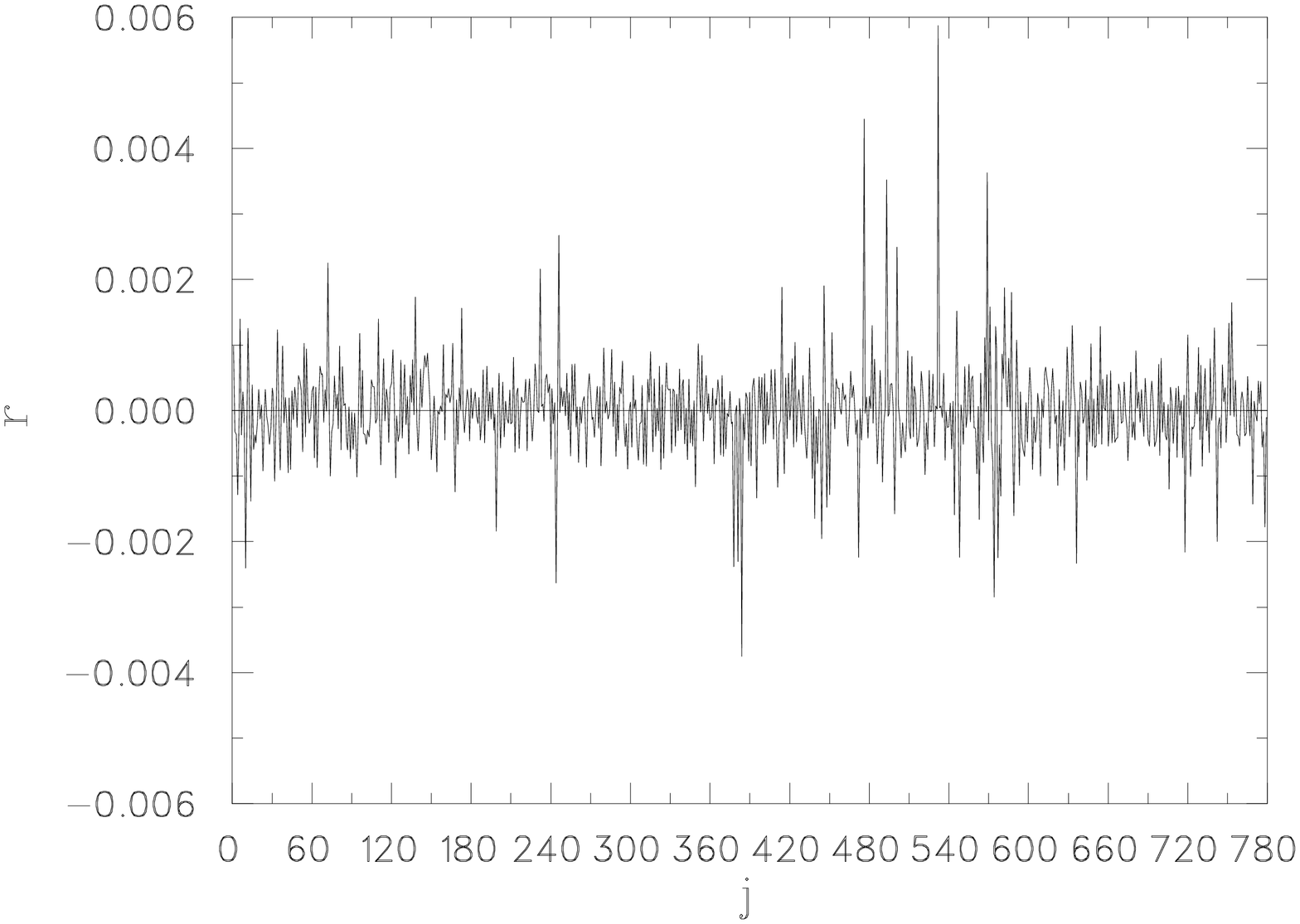}\hspace{3mm}
\includegraphics[angle=90,width=66mm]{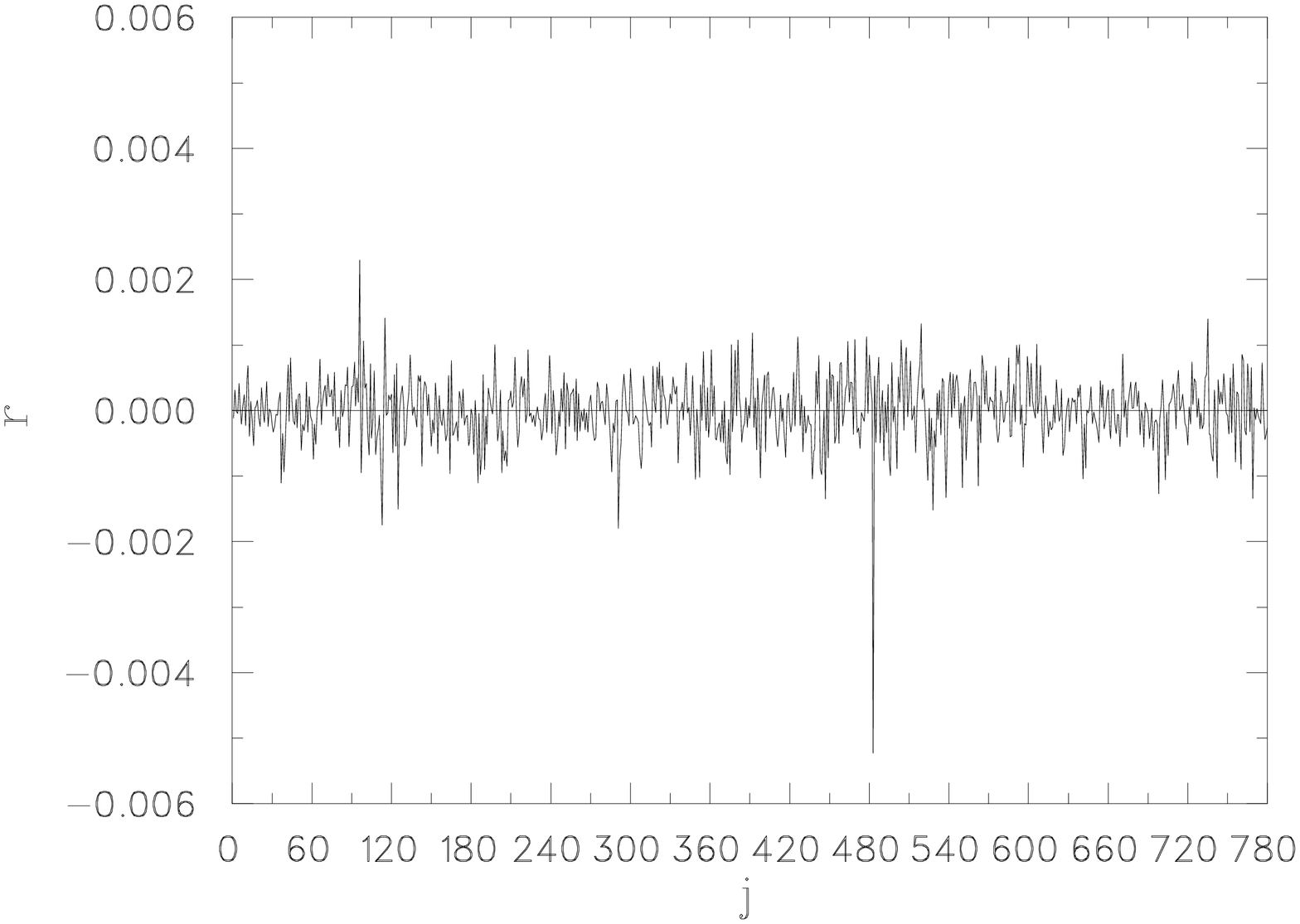}\vspace{4mm}\\
\rule{2mm}{0mm}
\includegraphics[angle=90,width=63mm]{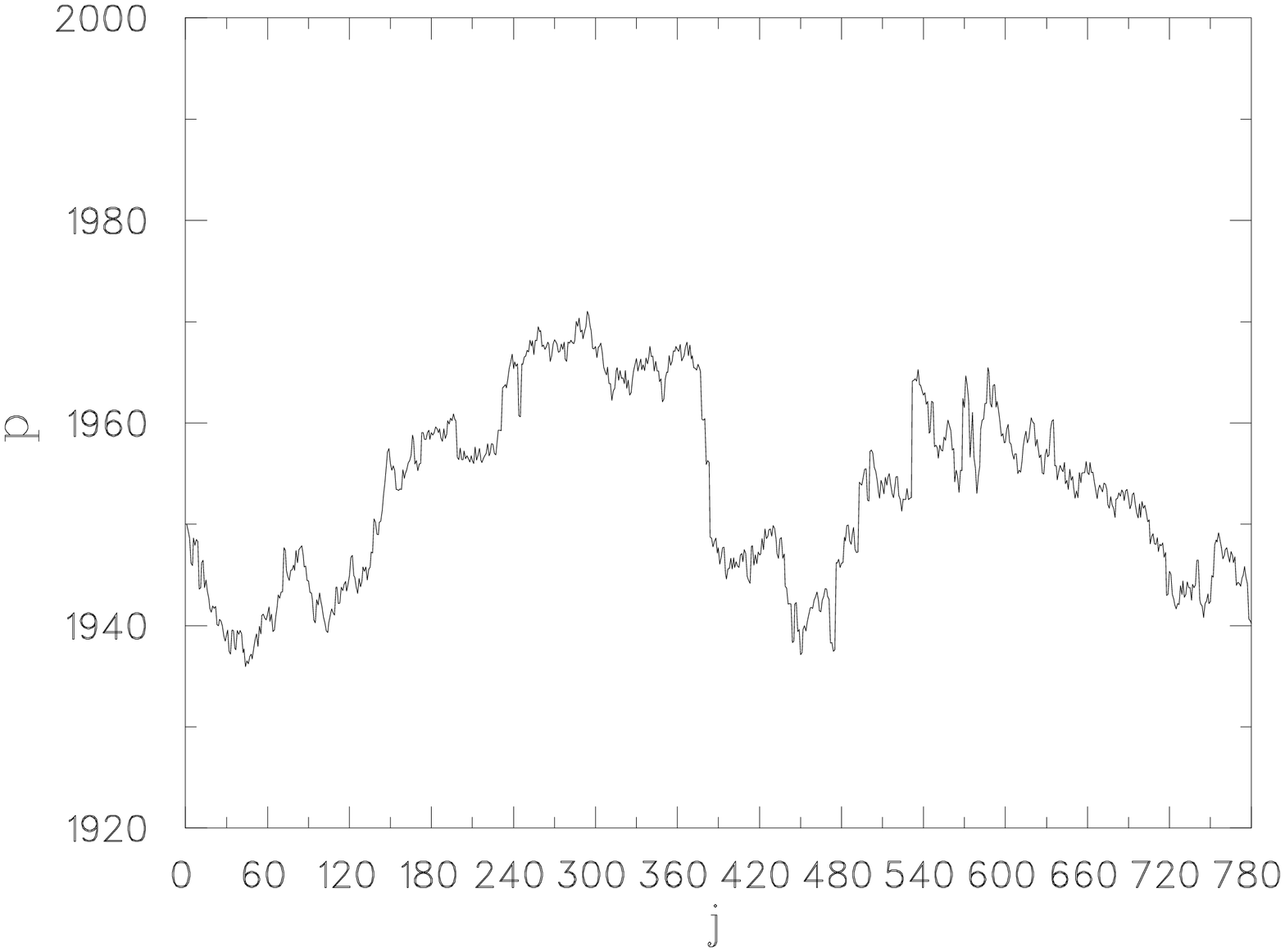}\hspace{6mm}
\includegraphics[angle=90,width=63mm]{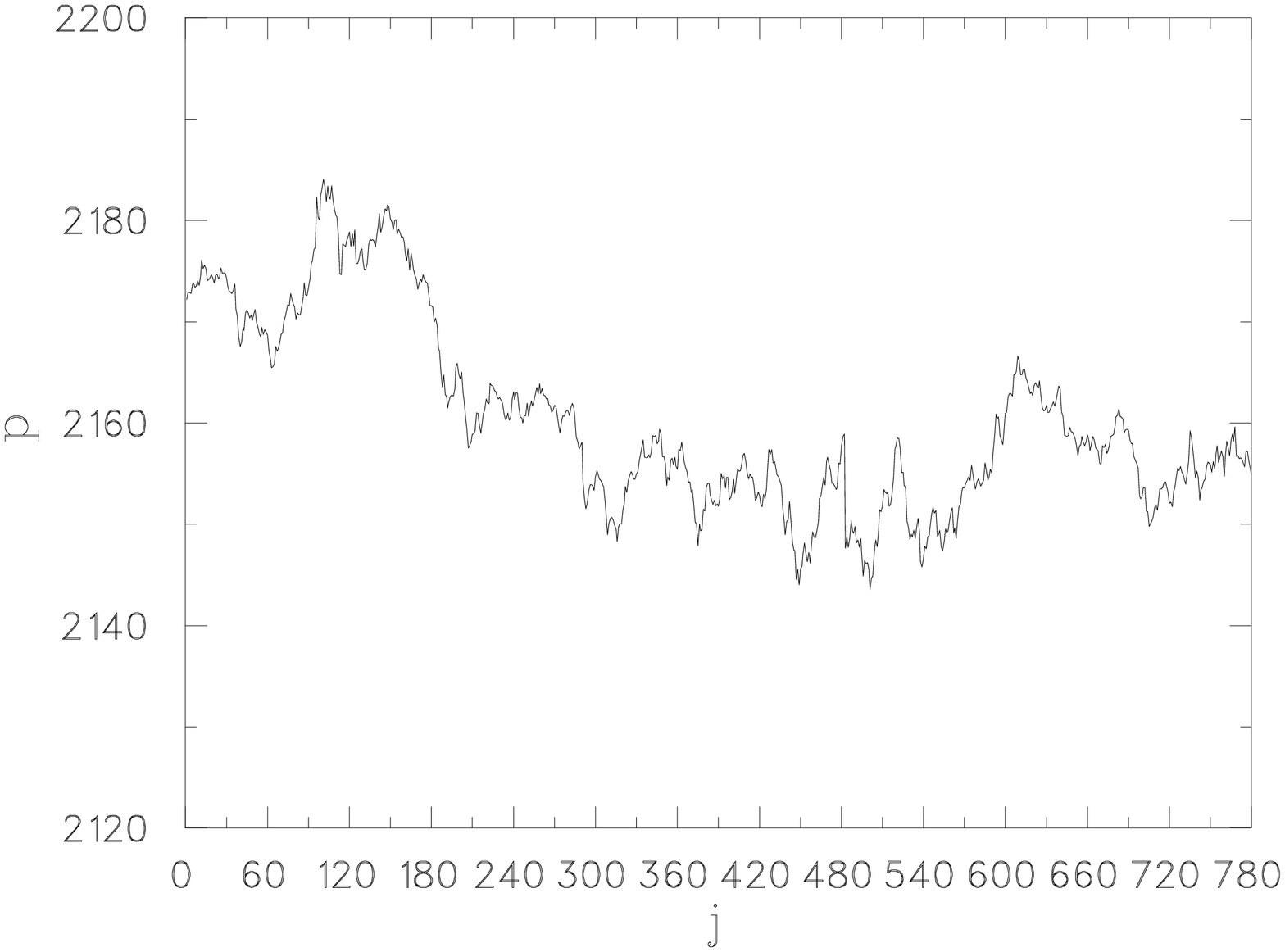}\vspace{4mm}\\
\rule{3mm}{0mm}
\includegraphics[angle=90,width=62mm]{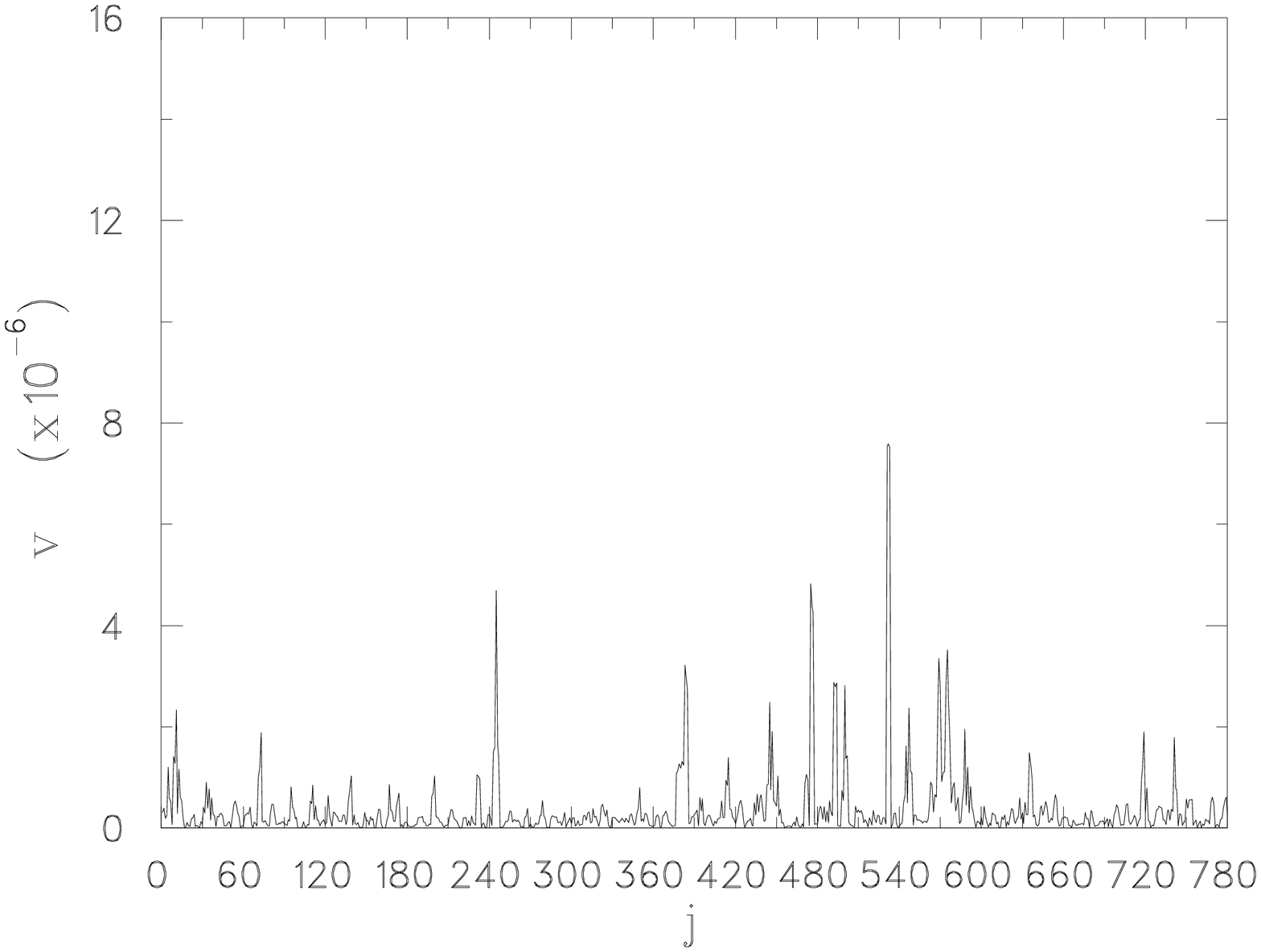}\hspace{7mm}
\includegraphics[angle=90,width=62mm]{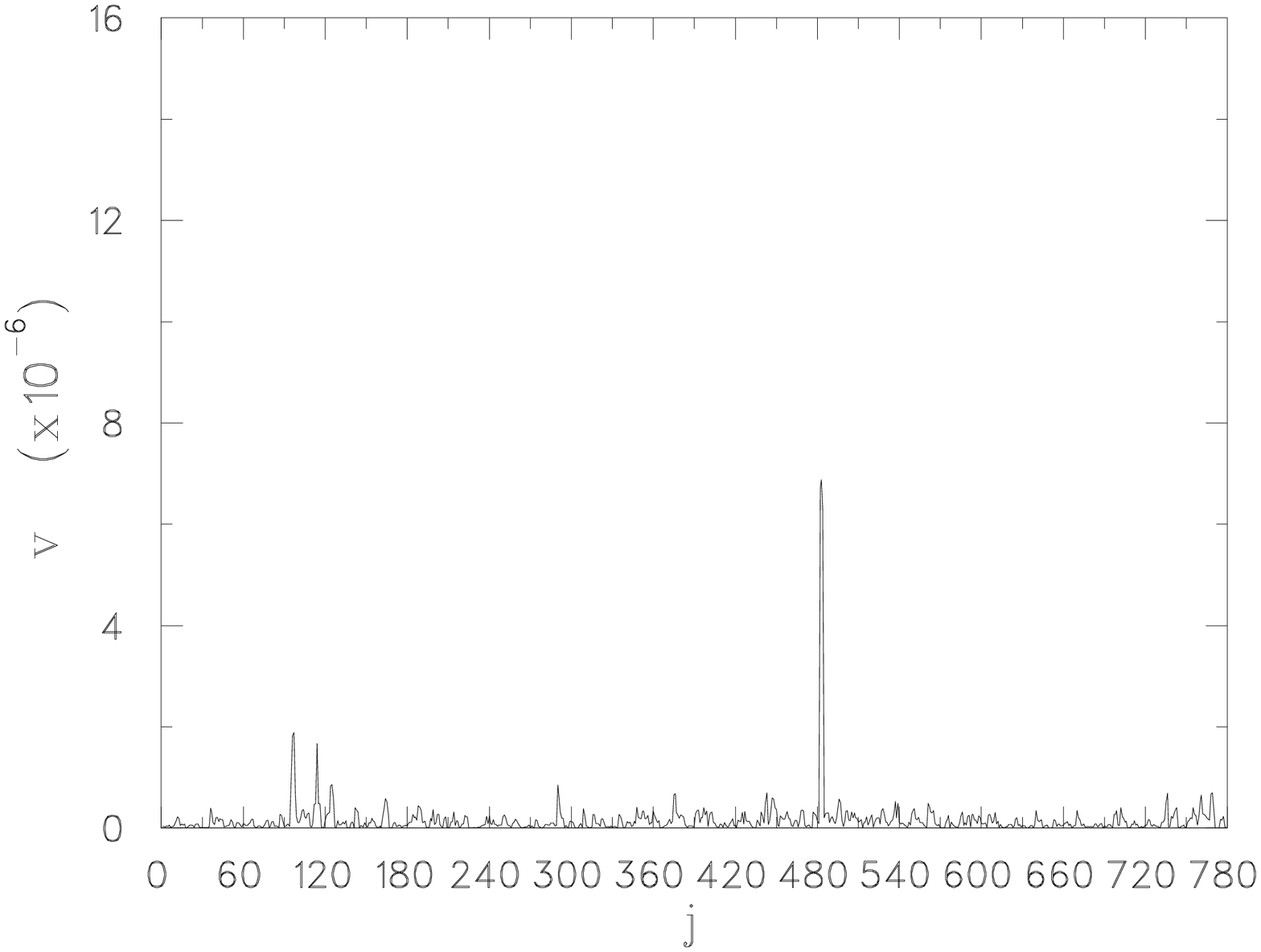}
\caption{\label{fig:flh3}Set 3: See caption of Fig.~\protect\ref{fig:flh1}.}
\end{figure}

\begin{figure}[p]
\includegraphics[angle=90,width=66mm]{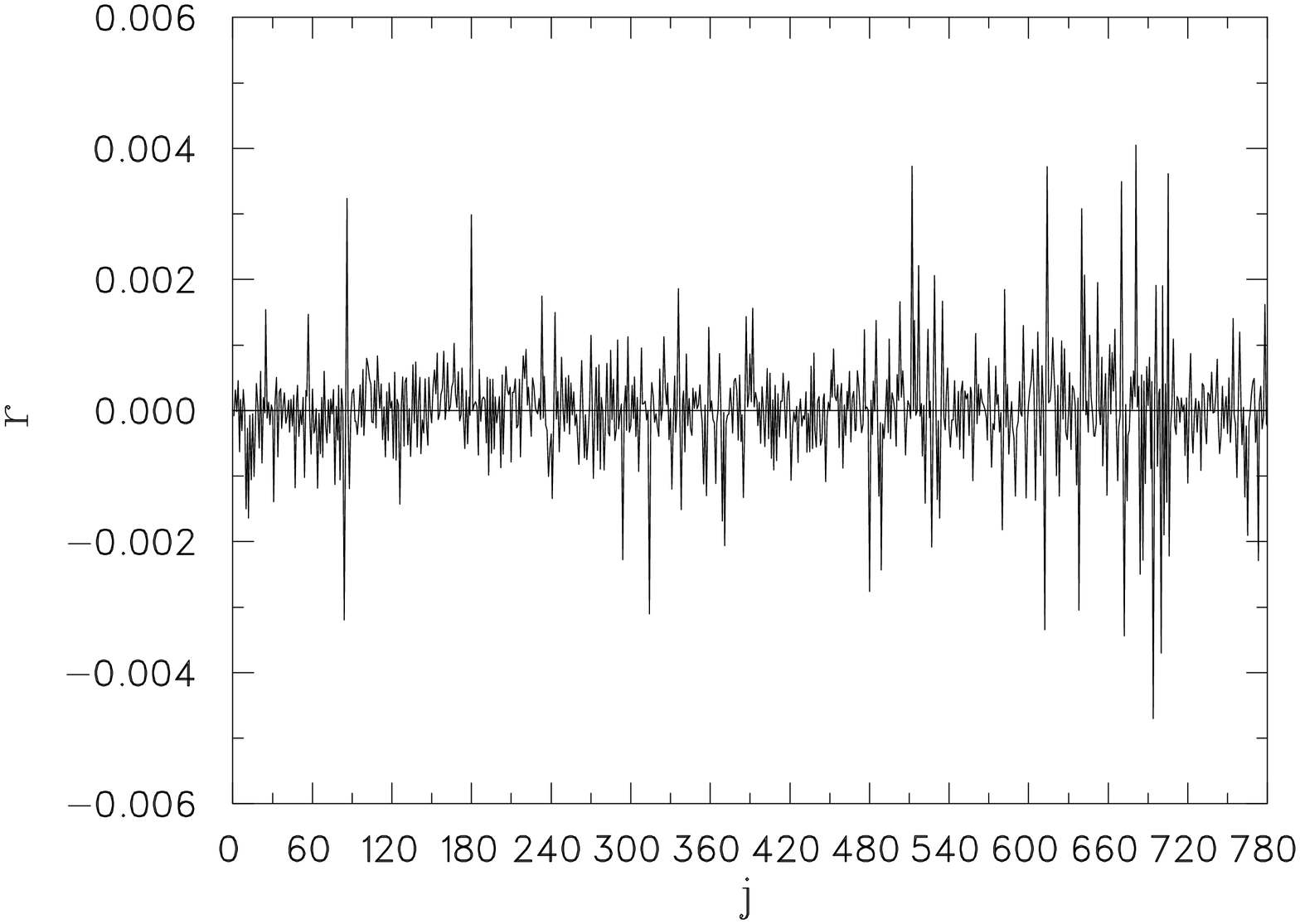}\hspace{3mm}
\includegraphics[angle=90,width=66mm]{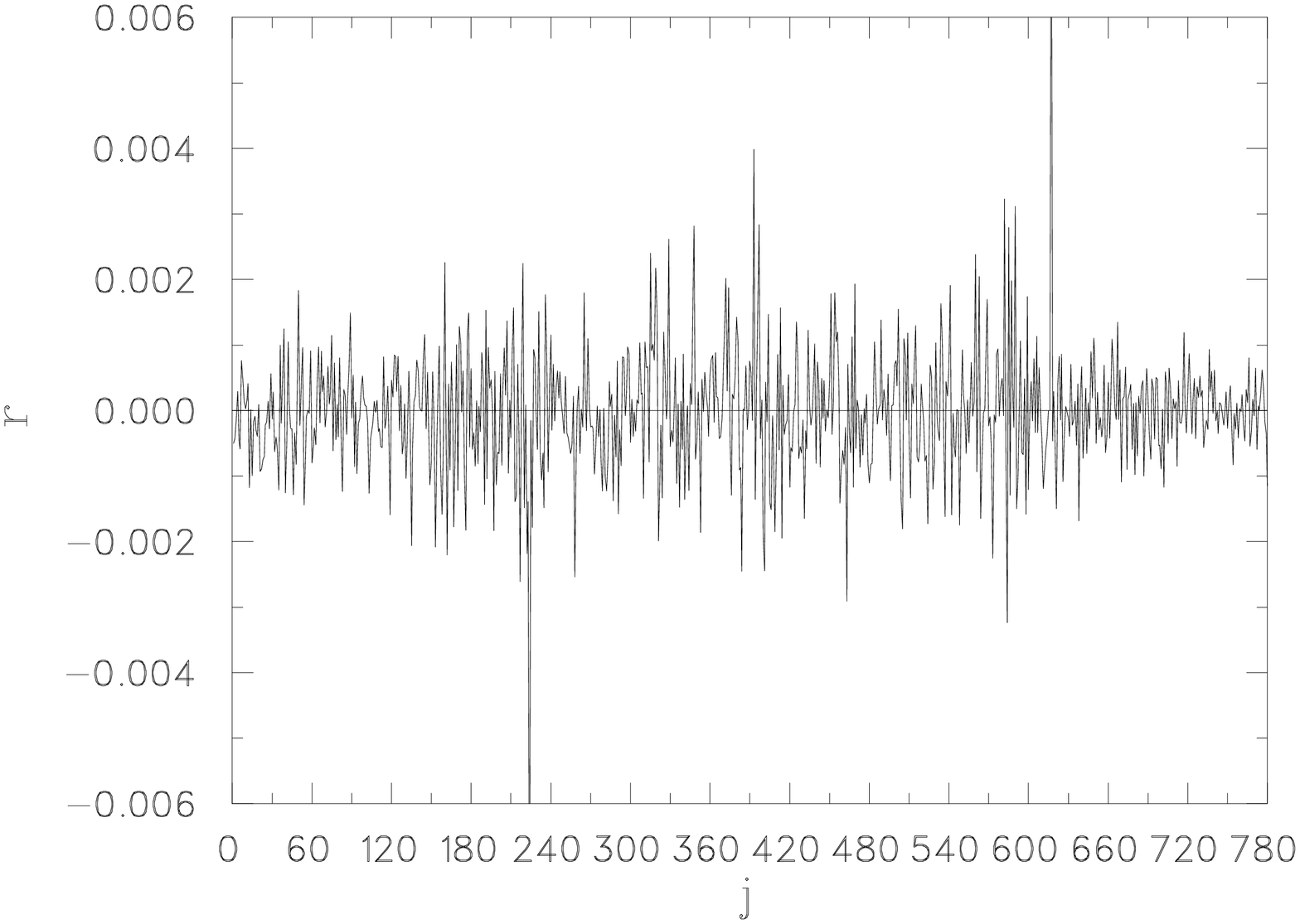}\vspace{4mm}\\
\rule{2mm}{0mm}
\includegraphics[angle=90,width=63mm]{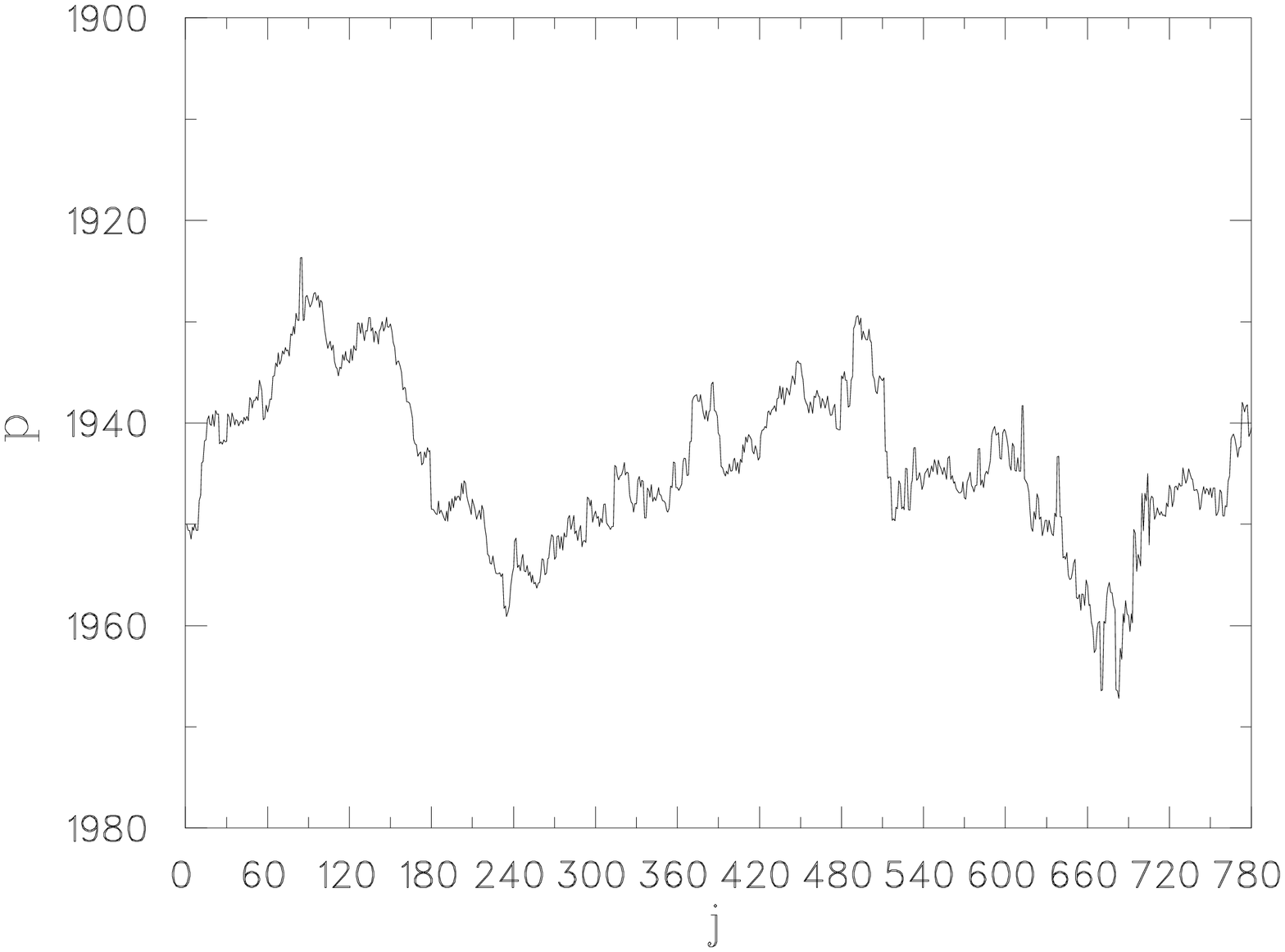}\hspace{6mm}
\includegraphics[angle=90,width=63mm]{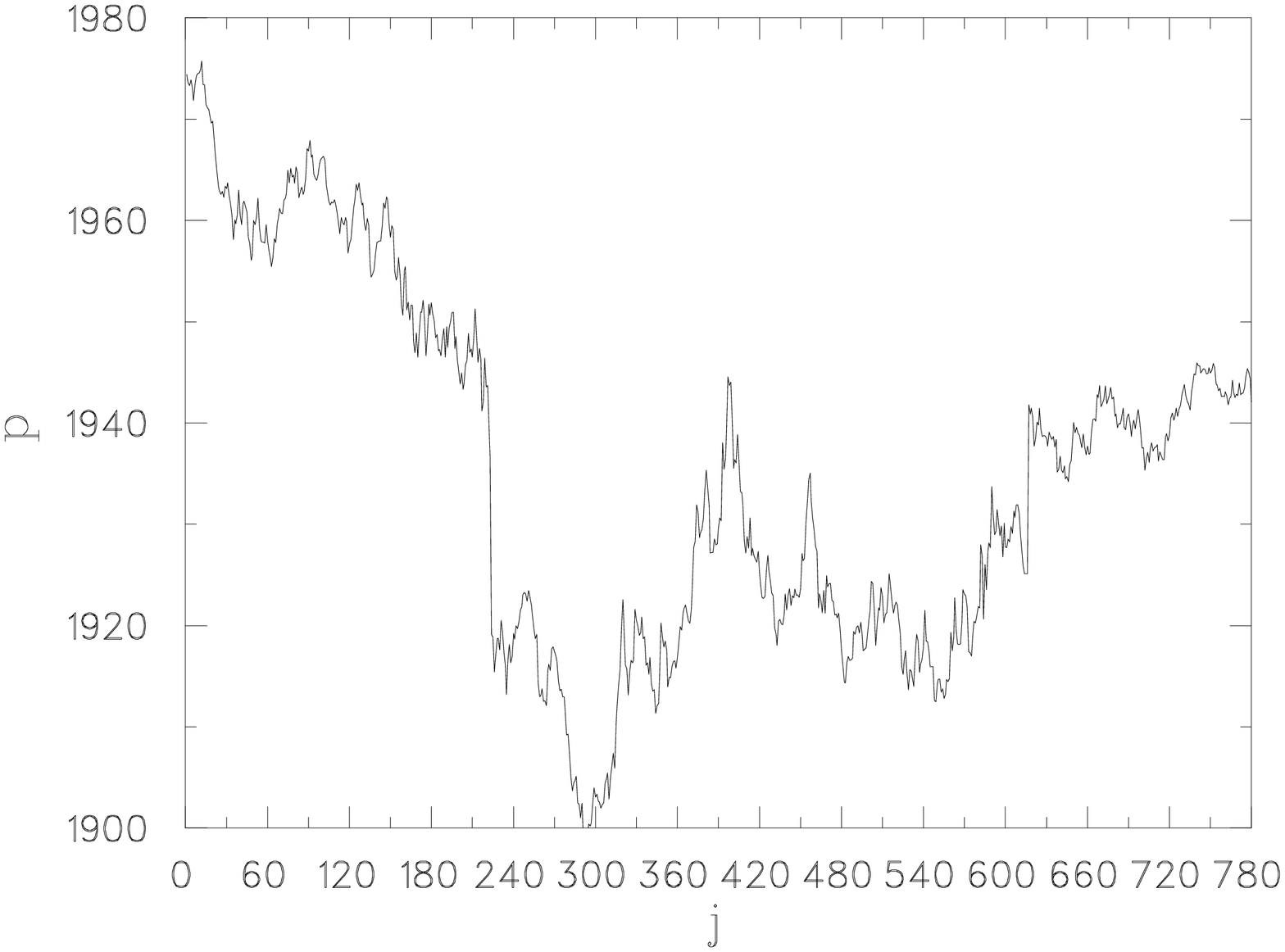}\vspace{4mm}\\
\rule{3mm}{0mm}
\includegraphics[angle=90,width=62mm]{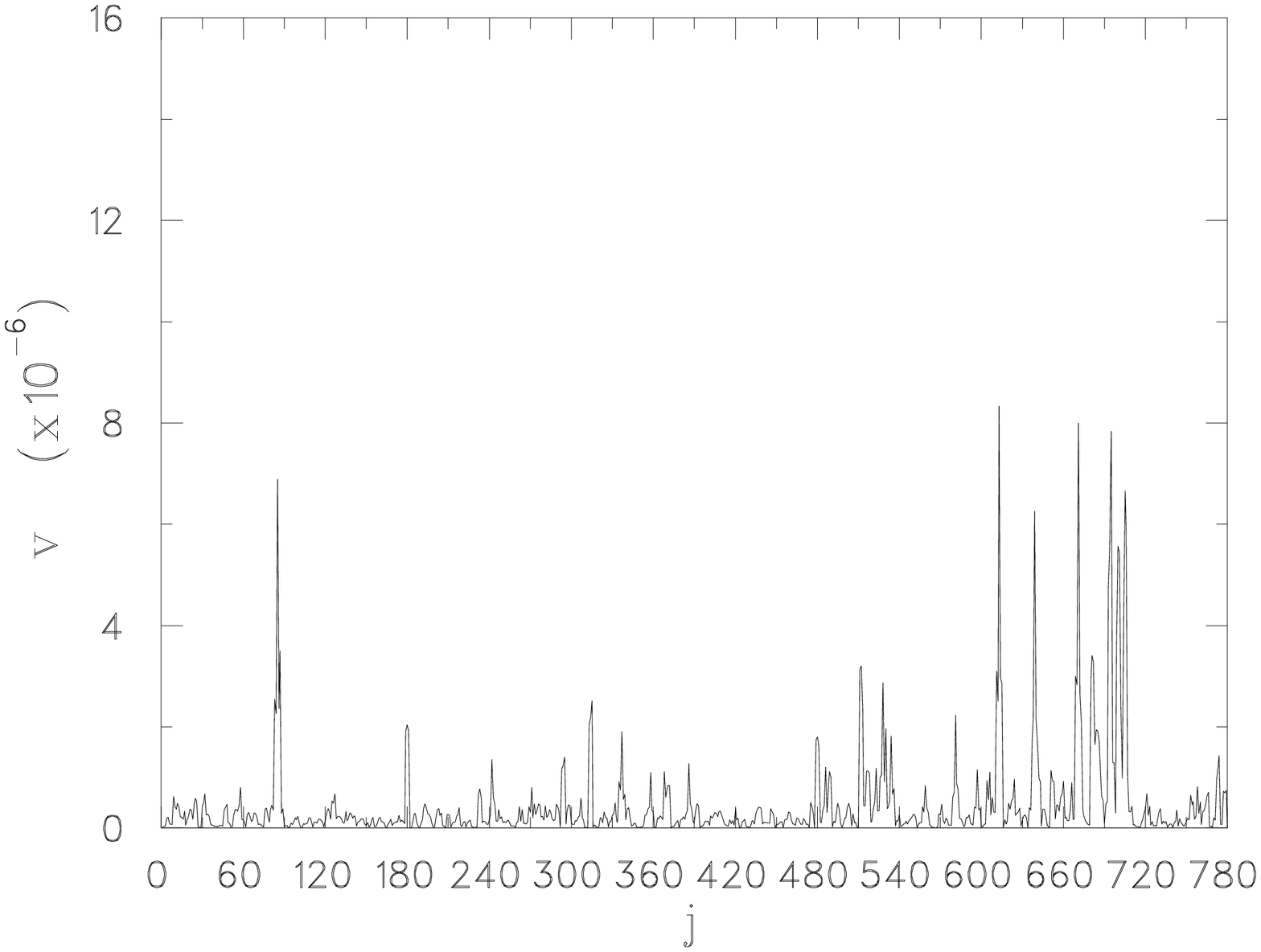}\hspace{7mm}
\includegraphics[angle=90,width=62mm]{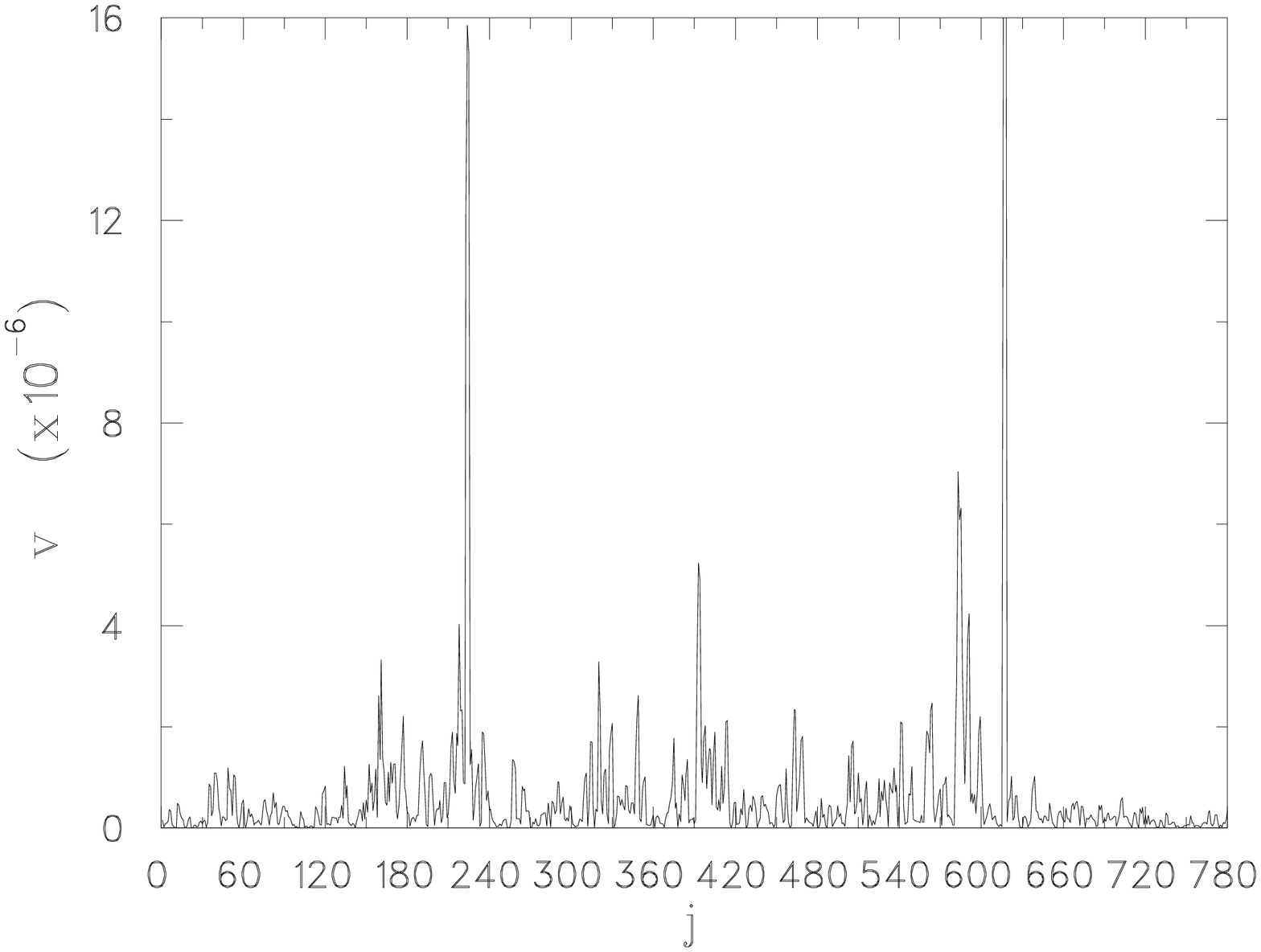}
\caption{\label{fig:flh4}Set 4: See caption of Fig.~\protect\ref{fig:flh1}.}
\end{figure}

In Figs.~\ref{fig:flh1}--\ref{fig:flh4} we show four sets of sample time series from
the lattice and selected historical data \cite{Finam:2008} from the NASDAQ index.
The latter are included to
demonstrate that the lattice model has the `distinctive air' of a real market.
It goes without saying that a statistical model is at a loss predicting time series,
nor can one expect that a single model will describe subtle details of stochastic
features of every market. What we are looking for has been very eloquently layed out
in \cite{Mandelbrot:1997} by Mandelbrot. In the Introduction we read:
``It is worth noting that fully fleshed-out
and detailed pictures ... put a heavy premium on the ability of the eye to recognize
patterns that existing analytic techniques were not designed to identify or assess.''
It is in this spirit that Figs.~\ref{fig:flh1}--\ref{fig:flh4} are presented.

The NASDAQ data from \cite{Finam:2008} are at minute intervals,
which naively translates to $\Delta=60{\rm s}$ for the parameter introduced
at the beginning of Sect.~\ref{sec:model}.
There are discontinuities by end-of-the-day and over-the-weekend interruptions
in the time series. Ignoring this, we have randomly picked time series of length $n$
from the historical data. However, the situation is more complicated.
Assuming that the time series has fractal geometry, and thus is devoid of a scale,
the choice of any $\Delta$ would be equally valid. We are not in a position to
pursue this issue here, but will rather live with the above naive choice,  
adopting the scaling argument.

As for Figs.~\ref{fig:flh1}--\ref{fig:flh4} the three panels on the left column
display, top to bottom, for $j=0\ldots n$ the returns $r_j$, the prices $p_j$
and the volatility $v_j$ of returns derived from the lattice model.
While the returns come directly from the simulation,
the price time series is obtained by integrating (\ref{eq8})
\begin{equation}
\Phi(t)=\Phi(t_0)\exp[\frac{1}{\Delta}\int_{t_0}^{t}dt^\prime r(t^\prime)]\,.
\label{eq12}\end{equation}
The discrete version of (\ref{eq12}) is equivalent to rewriting (\ref{eq1}) as
a recursion
\begin{equation}
p_j=p_{j-1}\exp(r_j)
\label{eq13}\end{equation}
with $p_j=\Phi_j C$ and initial condition $p_0$.
Finally, the time series of the variance, or volatility in financial terms,
is computed from just three time slices through
\begin{equation}
v_j=\frac{1}{3}\sum_{j^\prime=j-1}^{j+1}(r_{j^\prime}-\bar{r})^2\quad\mbox{with}\quad
\bar{r}=\frac{1}{3}\sum_{j^\prime=j-1}^{j+1}r_{j^\prime}\,.
\label{eq14}\end{equation}

Note that $\Delta,p_0$ and $C$ are all trivial parameters (units, etc) in the sense that they
have no effect on the simulation. The same is true for the variance $w$ used to draw
random numbers, from $p(x)\propto\exp(-x^2/2w)$, while updating the lattice,
see (\ref{eq6}). 
Since the signal (\ref{eq4}) of the field configuration only involves a comparison
($\max$) of returns the effect of changing $w$ will be a rescaling of the field. Specifically
\begin{equation}
w\rightarrow \lambda w \quad\mbox{then}\quad r_j\rightarrow \sqrt{\lambda}\, r_j\,.
\label{eq15}\end{equation}
Therefore, doing an entire simulation at only one fixed, arbitrary, choice for $w$ is sufficient.
In this sense the model has no adjustable parameters.
Observables then scale accordingly, for example $v_j\rightarrow \lambda v_j$.
The price time series
(\ref{eq13}) behaves in a less straightforward, though well defined,
manner because the scaling factor appears with the argument of an exponential function.
The plots in Figs.~\ref{fig:flh1}--\ref{fig:flh4} were consistently produced with
$w=1,\, \lambda=2\times 10^{-5}$, and $p_0=1950$.
The resulting scales are similar to those of the selected historical NASDAQ data.
Aside from lateral shifts of the price viewing windows,
comparable scales in all panels of Figs.~\ref{fig:flh1}--\ref{fig:flh4} are identical.
 
\begin{table}[p]
\center
\begin{tabular}{ccccc} \hline
              &  Set 1 L                      &  Set 1 N                      \\ \hline
 $\alpha_{0}$ &  3.23E-08(1.01E-08)[3.214841] &  1.11E-08(5.35E-09)[2.079218] \\
 $\alpha_{1}$ &  0.043332(0.008898)[4.870008] &  0.130702(0.019696)[6.636112] \\
 $\beta_{1}$  &  0.891148(0.028338)[31.44721] &  0.859163(0.026981)[31.84381] \\ \hline
\end{tabular}
\caption{\label{tab:GARCH-1} GARCH fit parameters of Set 1, FIG.~\protect\ref{fig:flh1},
for the lattice L and historical N data.}
\end{table}

\begin{table}[p]
\center
\begin{tabular}{ccccc} \hline
              &  Set 2 L                      &  Set 2 N                      \\ \hline
 $\alpha_{0}$ &  4.14E-09(1.61E-09)[2.564338] &  4.20E-09(1.07E-09)[3.916649] \\
 $\alpha_{1}$ &  0.022450(0.005070)[4.428308] &  0.031982(0.007263)[4.403283] \\
 $\beta_{1}$  &  0.966177(0.007889)[122.4713] &  0.951740(0.010025)[94.93278] \\ \hline
\end{tabular}
\caption{\label{tab:GARCH-2} GARCH fit parameters of Set 2, FIG.~\protect\ref{fig:flh2},
for the lattice L and historical N data.}
\end{table}
  
\begin{table}[p]
\center
\begin{tabular}{ccccc} \hline
              &  Set 3 L                      &  Set 3 N                      \\ \hline
 $\alpha_{0}$ &  6.38E-09(1.46E-09)[4.360360] &  5.87E-09(2.02E-09)[2.907091] \\
 $\alpha_{1}$ &  0.015515(0.003301)[4.700237] &  0.022312(0.008688)[2.568121] \\
 $\beta_{1}$  &  0.972397(0.004485)[216.8057] &  0.956839(0.014609)[65.49589] \\ \hline
\end{tabular}
\caption{\label{tab:GARCH-3} GARCH fit parameters of Set 3, FIG.~\protect\ref{fig:flh3},
for the lattice L and historical N data.}
\end{table}
  
\begin{table}[p]
\center
\begin{tabular}{ccccc} \hline
              &  Set 4 L                      &  Set 4 N                      \\ \hline
 $\alpha_{0}$ &  2.76E-08(5.42E-09)[5.092728] &  3.73E-09(2.70E-09)[1.378933] \\
 $\alpha_{1}$ &  0.045674(0.007692)[5.937874] &  0.069064(0.005237)[13.18884] \\
 $\beta_{1}$  &  0.911153(0.013877)[65.65998] &  0.936347(0.003615)[259.0275] \\ \hline
\end{tabular}
\caption{\label{tab:GARCH-4} GARCH fit parameters of Set 4, FIG.~\protect\ref{fig:flh4},
for the lattice L and historical N data.}
\end{table}

The four data sets in Figs.~\ref{fig:flh1}--\ref{fig:flh4} were chosen somewhat randomly,
except that they were paired to point out common features appearing in both the lattice
and historical sets. For example, in Fig.~\ref{fig:flh1} there is a clearly visible time
interval around $j\approx 330$ where the historical returns data exhibits a region of
quenched volatility. In financial market data the phenomenon of volatility clustering
is well known and even exploited as an industry standard modeling
technique \cite{Engle:1982:ARCH}. Periods of small/large volatility are often separated
by price shocks that manifest themselves in large spikes of the volatility.
This is clearly a common feature of both the historical and the lattice data sets
of Fig.~\ref{fig:flh1}, best visible in the $v$ panels.
The price evolutions are similar, but this is accidental. The lattice model generates
up or down markets with equal probability. (Although this can easily be changed by
modifying the updating algorithm.) 

The remaining sets exhibit the same characteristics. In Set~2, Fig.~\ref{fig:flh2}, we
have paired lattice and historical data that share a small overall volatility.
This demonstrates that the lattice model is capable of producing quiescent market
periods as well. The price times series look strikingly similar, which is fallacious,
the statistical model is not capable of making predictions.
The subsequent Sets~3 and 4 provide additional evidence that the lattice model is
able to emulate stylized featured of real markets. Again, volatility clusters are
clearly present separated by spikes of various sizes.
Similarly, Set~4 exhibits fairly active volatility patterns
for both the lattice and the historical data.

\subsection{Financial market dynamics}

We now turn to the ability of our model to capture some of the well-known dynamics observed in 
financial markets. One of the most important characteristics of a financial time series is its 
volatility, and more importantly, how the volatility evolves over time. Most financial time series 
exhibit time-varying volatility clustering, which means that periods of large swings tend to be 
followed by periods of large swings, while periods of calm tend to be followed by periods of calm. 
These dynamics can be modeled by the Auto Regressive Conditional Heteroskedasticity (ARCH) model of 
Engle \cite{Engle:1982:ARCH} and by its generalized version, the Generalized Auto Regressive
Conditional Heteroskedasticity (GARCH) model of Bollerslev \cite{Bollerslev:1986:GARCH}.

The fact that these specifications imply that a large shock on average tends to be followed by 
another large shock means that the resulting distribution of returns will exhibit `fat tails' or 
higher-than-Gaussian probability masses in the extreme regions. Similarly, the fact that these 
specifications also imply that a small shock on average tends to be followed by another small shock 
means that the resulting distribution of returns will exhibit a higher-than-Gaussian probability 
mass around the origin, see Fig.~\ref{fig:hd}.
ARCH and GARCH models have grown into incredibly popular tools as they are 
able to replicate these salient features of financial returns distributions, namely more 
probability mass in the tails and around the center of the distribution than in the benchmark 
Gaussian case. Capturing these features of financial markets is crucial for many purposes including 
but not limited to: derivatives pricing, hedging, forecasting volatility, portfolio management, 
regulatory issues, value-at-risk, and so on.

The traditional ARCH specification expresses the current volatility level as a function of past 
squared shocks (about a mean or average return) at various lags. This implies that the volatility 
is not constant over time anymore but that it depends on how large recent deviations (positive or 
negative) from the mean have been. In the more general GARCH specification, the current volatility 
level is still a function of past squared shocks to the returns, but it is also a function of past 
lagged levels of itself, thus making the model even more flexible. The GARCH(p,q) specification for 
the volatility $\sigma^2_t$ at time $t$ can be described by the following equation:
\begin{equation}
\sigma^2_t = \alpha_0+\sum_{i=1}^{q}\alpha_i\epsilon_{t-i}^{2}+\sum_{i=1}^{p}\beta_i\sigma_{t-i}^{2}
\label{eq22}\end{equation}
where $q$ and $p$ are the maximum lags allowed by the model for past shocks and volatility levels 
respectively, and $\epsilon_{t-i}$ is the return shock (about the mean) at lag $i$.

One can obviously allow $q$ and $p$ to be as large as one wants, but a GARCH(1,1) model in practice 
turns out to be surprisingly flexible. Moreover, overfitting is often the recipe for poor 
out-of-sample performance, and parsimonious models often end up defeating more complex ones when 
tested outside of the in-sample period.
Hansen and Lunde \cite{Lunde+Hansen:2005} compare 330 (G)ARCH-type models in 
terms of their ability to describe the conditional volatility of exchange rate and IBM return data, 
and find no evidence that a GARCH(1,1) is outperformed by more sophisticated models in the 
analysis. Therefore we choose to focus on the parsimonious, yet very apt, GARCH(1,1) model for 
purposes of comparing the dynamics of time-varying volatility in our lattice-generated returns and 
in NASDAQ historical returns. Our model can thus be written as
\begin{equation}
\sigma^2_t = \alpha_0+\alpha_1\epsilon_{t-1}^{2}+\beta_1\sigma_{t-1}^{2} \,.
\label{eq23}\end{equation}
	 
The goal here is twofold. First, it is to investigate whether our lattice model is able to produce 
returns volatility dynamics displaying some form, if any, of ARCH/GARCH effects. Second, it is to examine 
whether our lattice model is able to produce returns volatility dynamics that are rather consistent 
with those of NASDAQ historical returns. In figures \ref{fig:flh1} through \ref{fig:flh4}
for each set L and N (lattice and 
NASDAQ returns), we fit a GARCH(1,1) model onto the returns data through a Levenberg-Marquardt 
optimization algorithm \cite{levenberg:1944,marquardt:431}
and report the results in tables \ref{tab:GARCH-1} through \ref{tab:GARCH-4}.

Before getting to the tables, one may first notice that
figures \ref{fig:flh1} through \ref{fig:flh4} display 
similar-looking sets of charts when one compares the lattice-generated graphs with the NASDAQ 
historical ones. The top portion represents returns (gains or losses) over time, and one can 
readily see that there are clusters of volatility in each, indicating that the model seems to be 
capable of capturing them. The middle portion simply represents the evolution of the asset value 
over time. Visually, the price dynamics generated by the lattice model appear `plausible' as those 
of a financial market such as the NASDAQ. Finally, the bottom plot represents the evolution of the 
volatility over time, for both the lattice and the NASDAQ data. Here again, the volatility patterns 
generated by the lattice model seem credible as those of financial markets, displaying a variety of 
spikes of various sizes and frequencies.

Underneath the plots, each table displays the estimated 
parameters $\alpha_0$, $\alpha_1$ and $\beta_1$,
followed by their standard errors in parenthesis and their t-statistics,
in square brackets, for both the lattice and the NASDAQ returns data.
Except for the NASDAQ parameter $\alpha_0$ in Tab.~\ref{tab:GARCH-4}, every parameter is
statistically significant at the 5\% significance level, and most 
parameters are even statistically significant at the 1\% significance level. This indicates that 
both the NASDAQ returns and our lattice-generated returns display ARCH/GARCH behavior in 
volatility. Moreover, and more remarkably, the estimated parameters coming from our 
lattice-generated returns are extremely close, especially in magnitude, to the estimated parameters 
coming from the NASDAQ historical returns. For instance, in Tab.~\ref{tab:GARCH-2},
$\alpha_0$, $\alpha_1$ and $\beta_1$ are estimated to 
be $4.14\times 10^{-9}$, $0.022450$ and $0.966177$ respectively for the lattice-generated model,
while they are 
$4.20\times 10^{-9}$, $0.031982$ and $0.951740$ respectively for the NASDAQ historical returns.
Although this 
does not indicate forecasting abilities on the part of the lattice model, it does show that it is 
able to reproduce, with precision, important financial markets volatility dynamics and thus has the 
potential to provide future insights on the inner mechanics of such markets.

\section{\label{sec:conclude}Summary and conclusion}

By their very nature, historical market data constitute only instances of some
stochastic process. 
It is thus desirable to have available a stochastic model of a financial market.
As a result, such a model grants
complete access to the market's
stochastic features through the act of drawing multiple instances, or ensembles.
This allows the possibility of a detailed investigation of market dynamics and
the features that define them.

We have studied the properties of a stochastic lattice model which by design
derives its features from next-nearest-neighbor interactions of microscopic
entities that live on a linear chain in discrete time.
The model is realized as a lattice field theory with field components being
interpreted as market returns, and is subject to
computer simulation with an updating algorithm inspired by the evolution model
by Bak, Tang and Wiesenfeld \cite{PhysRevLett.59.381}
that drives the lattice field into a self-organized critical state.
We present evidence, including power law behavior, that the critical state is indeed
achieved. Its presence is the salient feature of the model.

We compute time series of market returns, prices, volatilities, and returns
frequency distributions, all of which are remarkably
consistent with historical market data for the NASDAQ stock index.
In particular the `fat tails' feature of the returns distribution comes out
effortlessly. It is worth noting that, aside from (trivial) units,
the lattice model has no adjustable parameters.
Perhaps most remarkable is the observation that standard financial industry
analysis tools, in our case the GARCH(1,1) model \cite{Bollerslev:1986:GARCH},
produce fits that make the lattice model time series almost indistinguishable from
real financial market data.

It seems that the most important conclusion derived from our study is that
self-organized criticality, being a fundamental driving force for the model,
is also key to characterizing real world financial markets.
To say the least, self-organized criticality should be an important component of intrinsic market dynamics,
allowing us to model financial instruments using
lattice methods which may hopefully be competitive with current industry practice.

Of course, self-organization is surely not the only mechanism that drives financial
market dynamics. Among other things, trading occurs within a background of arbitrage
opportunities \cite{Ilinski:2001:PF}.
We are looking forward to combining past studies focused on arbitrage \cite{Dupoyet2010107}
with the present model in order to develop an even more realistic stochastic market model.

\end{document}